\documentclass[twocolumn,preprintnumbers,superscriptaddress,endnote,nofootinbib,aps,prd]{revtex4-1}

\pdfoutput=1

\usepackage[hypertex]{hyperref}

\usepackage{graphicx}
\usepackage{dcolumn}   
\usepackage{amssymb}   
\usepackage{amsmath}

\usepackage{colordvi}
\usepackage{epsfig}
\usepackage{latexsym}

\newcommand{ \centeron }[2]{{\setbox0=\hbox{#1}\setbox1=\hbox{#2}\ifdim
                             \wd1>\wd0\kern.5\wd1\kern-.5\wd0\fi \copy0
                             \kern-.5\wd0\kern-.5\wd1\copy1\ifdim\wd0>\wd1
                             \kern.5\wd0\kern-.5\wd1\fi}}
\newcommand{ \ltap }{\>\centeron{\raise.35ex\hbox{$<$}}
                     {\lower.65ex\hbox{$\sim$}}\>}
\newcommand{ \gtap }{\>\centeron{\raise.35ex\hbox{$>$}}
                     {\lower.65ex\hbox{$\sim$}}\>}

\newcommand{ \slashchar }[1]{\setbox0=\hbox{$#1$}   
   \dimen0=\wd0                                     
   \setbox1=\hbox{/} \dimen1=\wd1                   
   \ifdim\dimen0>\dimen1                            
      \rlap{\hbox to \dimen0{\hfil/\hfil}}          
      #1                                            
   \else                                            
      \rlap{\hbox to \dimen1{\hfil$#1$\hfil}}       
      /                                             
   \fi}                                             %

\newcommand{\ETmiss}{\slashchar{E}_T}

\newcommand{\gev}{\text{GeV}}
\newcommand{\tev}{\text{TeV}}

\newcommand{\lsim}{\lesssim}

\newcommand{\ra}{\rightarrow}

\newcommand{\HTjets}{H_T^{\mathrm{jets}}}

\begin{document}

\pagestyle{plain}

\preprint{FERMILAB-PUB-10-190-T}

\title{Discovering Higgs Bosons of the MSSM using Jet Substructure}

\author{Graham D. Kribs}
\affiliation{Department of Physics, University of Oregon,
             Eugene, OR 97403}

\author{Adam Martin}
\affiliation{Theoretical Physics Department, Fermilab, Batavia, IL 60510}

\author{Tuhin S. Roy}
\affiliation{Department of Physics, University of Oregon,
             Eugene, OR 97403}

\author{Michael Spannowsky}
\affiliation{Department of Physics, University of Oregon,
             Eugene, OR 97403}


\begin{abstract}

We present a qualitatively new approach to discover Higgs bosons 
of the MSSM at the LHC using jet substructure techniques 
applied to boosted Higgs decays.  These techniques are
ideally suited to the MSSM, since the lightest Higgs boson 
overwhelmingly decays to $b\bar{b}$ throughout the entire 
parameter space, while the heavier neutral Higgs bosons, 
if light enough to be produced in a cascade, 
also predominantly decay to $b\bar{b}$.  
The Higgs production we consider arises from superpartner 
production where superpartners cascade decay into Higgs bosons.  
We study this mode of Higgs production for several superpartner
hierarchies:  $m_{\tilde{q}} , m_{\tilde g} > m_{\tilde{W},\tilde{B}} > m_h + \mu$;
$m_{\tilde{q}} , m_{\tilde g} > m_{\tilde{W},\tilde{B}} > m_{h,H,A} + \mu$;
and $m_{\tilde{q}} , m_{\tilde g} > m_{\tilde{W}} > m_h + \mu$ with 
$m_{\tilde{B}} \simeq \mu$.
In these cascades, the Higgs bosons are boosted, 
with $p_T > 200$~GeV a large fraction of the time.
Since Higgses appear in cascades originating from squarks and/or
gluinos, the cross section for {\it events with at least one
Higgs} can be the same order as squark/gluino production.
Given $10$~fb$^{-1}$ of 14 TeV LHC data, with 
$m_{\tilde{q}} \lsim 1$~TeV, and one of the above superpartner 
mass hierarchies, our estimate of $S/\sqrt{B}$ of the Higgs signal 
is sufficiently high that the $b\bar{b}$ mode can become 
\emph{the} discovery mode of the lightest Higgs boson of the MSSM.

\end{abstract}
\maketitle

\section{Introduction}

Uncovering the origin of electroweak symmetry breaking is
of the utmost importance for the LHC.  
If the world is supersymmetric -- in the form of the
minimal supersymmetric standard model (MSSM) --
electroweak symmetry breaking is accomplished through a 
supersymmetrized two-Higgs-doublet model, with couplings 
and interactions set or restricted by supersymmetry.

It is well known that imposing the proper electroweak symmetry 
breaking minimum leaves two undetermined parameters in the Higgs sector
at tree-level:  
the ratio of the Higgs scalar vevs, $\tan\beta$, and the mass of the 
CP-odd scalar, $m_A$.  Radiative corrections dominantly affect
the lightest Higgs mass, raising it from the ruled-out 
tree-level value $m_h = M_Z$ up to about $125$~GeV for 
stop masses and mixings that do not exceed $1$~TeV \cite{Carena:2000dp}. 
Decay rates of the Higgs bosons can also be computed largely
independently of the details of the superpartner sector
(so long as decays into superpartners are either kinematically
forbidden or rarely occur).  The Higgs sector can thus seemingly be
approximately parametrized by $m_h, m_A, \tan\beta$.  

This has reinforced the simplification that the Higgs scalar 
sector can be searched for, discovered, or ruled out in isolation from 
the remainder of the model~\cite{mssmlist}.  A casual glance at the ATLAS or CMS TDRs~\cite{deRoeck:942733, Aad:2009wy} demonstrates this canonical view, in which discovery potential for
the Higgs sector is plotted in the $m_A$-$\tan\beta$ plane (with some 
additional restriction on $m_h$ larger than the LEP II bound).  
The Achilles heal of this simplification is the assumption that
the most promising production channels of the Higgs bosons are 
largely the same ones as in the Standard Model (SM).

We demonstrate there is potentially a \emph{much superior} way to 
discover Higgs bosons in the MSSM -- superpartner production with 
superpartners that cascade decay into Higgs bosons.  
Higgs bosons from superpartner cascades is not a new idea, see e.g.~\cite{Baer:1992ef,Hinchliffe:1996iu,Datta:2001qs,Datta:2003iz,Bandyopadhyay:2008fp,Huitu:2008sa,Bandyopadhyay:2008sd,Fowler:2009ay}, 
but our method for 
finding and identifying Higgs bosons within the supersymmetric event sample 
is qualitatively new.  We exploit recently developed jet substructure
techniques \cite{Butterworth:2008iy} with modifications that we
presented in Ref.~\cite{Kribs:2009yh} 
to isolate the boosted Higgs-to-$b\bar{b}$ signal 
from the Standard Model and supersymmetric backgrounds.
The existence of a large supersymmetric cascade-to-Higgs rate 
requires relatively mild assumptions about the superpartner 
mass hierarchy. 

The notion to find and study supersymmetric signals through the 
hadronic decays of gauge bosons, as well as the lightest Higgs boson, 
was pointed out in an early use of jet substructure in 
Ref.~\cite{Butterworth:2007ke}.  
There, however, the motivation was \emph{not} to find the Higgs, 
but instead to recover the superpartner mass spectrum using
a kinematical edge analysis.  

In our previous paper \cite{Kribs:2009yh}, we pointed out that a signal 
of the Higgs boson itself can often be much more easily found within 
new physics, since the new physics can have a larger 
production cross section and larger fraction of boosted Higgs bosons.  
But identifying the Higgs boson in processes that invariably 
have busier final states with more jets (and potentially more 
hard $b$-jets) required modest improvements to the BDRS algorithm.  
This is not unlike the situation faced by Ref.~\cite{Plehn:2009rk}
in proposing a method to extract the Higgs signal from 
$t\bar{t}h$ production.

The \emph{commonalities} between our previous work, Ref.~\cite{Kribs:2009yh}, 
and this paper are:
\begin{enumerate}
\item We seek two-body decays of a Higgs boson into $b\bar{b}$.  
\item We use the same jet substructure algorithm to 
      extract this Higgs signal. 
\item The Study Points in this paper are pure MSSM\@.
\item We apply fairly aggressive cuts to reduce the backgrounds 
from standard model processes.
\end{enumerate}

The main \emph{differences} between Ref.~\cite{Kribs:2009yh} 
and this paper are:
\begin{enumerate}
\item The LSP of the MSSM-based Study Points in this paper is a 
neutralino\footnote{It could also be a ``neutralino-equivalent'',
where the neutralino is a NLSP and the gravitino is LSP, 
but the lifetime of the NLSP is sufficiently long that its
decays are not observed within the collider detectors.}.  
Our previous work, instead considered study points with 
a gravitino LSP and a promptly
decaying Higgsino NLSP\@.
\item The new physics signal is large missing energy, 
with characteristically large $\HTjets$.  (Our previous work, 
instead considered the new physics signal to be one hard 
$\gamma$ plus missing energy.)  This means that while the 
LHC will have evidence for new physics
in channels involving large missing energy, it quite unlikely 
that the new physics signals can be readily identified with 
specific processes or decays (or models, for that matter).  
\item In this paper, we also consider the detection of 
$H$ and $A$ states decaying to $b\bar{b}$, using jet substructure 
techniques.  We demonstrate that for lighter $m_A \lsim 200$~GeV, 
it is possible to uncover evidence both \emph{both} 
$h$ and $H/A$ with just 10 fb$^{-1}$ of data.
\end{enumerate}

The organization of the paper is as follows:
In Sec.~\ref{sec:goldstone} we explain how Higgs bosons 
can be produced from specific kinds of two-body superpartner decays.  
The main emphasis is on the qualitative features of gaugino and 
Higgsino interactions, so as to present a very clear picture of 
what superpartner hierarchies provide the most promising source 
of Higgs bosons, and how the large rates can be easily understood. 
In Sec.~\ref{sec:cascade} we consider the typically largest 
production source of heavy gauginos, namely, squark production and decay into 
gauginos.  We clearly demarcate which superpartners decay into
which gauginos, so that further studies can be guided by
these basic observations.  We then consider, for specific 
hierarchies, the prospect of finding a boosted Higgs in one of
these supersymmetric cascades.  We show that in a considerable region 
of the supersymmetric parameter space, 
as many as one in four typical decay chains originating in a squark and
ending in the LSP can contain a significantly boosted Higgs boson.
In Sec.~\ref{sec:relic} we show that the supersymmetric parameter
space we consider naturally satisfies the upper bounds on the thermal 
relic density.  Moreover, we demonstrate how simple changes in the
gaugino mass hierarchy (lowering $M_1$) can result in matching
the cosmological density, but without significantly affecting 
the Higgs boson signal.
In Sec.~\ref{sec:substructure} we then discuss the techniques
and algorithm to find Higgs decay using jet substructure.
We compare and contrast our methodology with what has been
used before for Standard Model production of a Standard Model 
Higgs boson.  In Sec.~\ref{sec:results} we present a series of
Study Points that demonstrate the effectiveness of our algorithm in 
finding one or more Higgs bosons of the MSSM\@.  The series of
plots of candidate resonance jet mass are the main results of this
paper -- demonstrating that the $b\bar{b}$ mode could well be
\emph{the} discovery mode of Higgs bosons at the LHC\@.
Finally, in Sec.~\ref{sec:discussion} we conclude with a 
discussion of our results.

\section{Higgs from Superpartner Decay:  ``Goldstone region''}
\label{sec:goldstone}

The main focus of the paper is on Higgs bosons that arise
from the two-body decays of neutralinos and charginos, 
\begin{eqnarray}
\chi^0_i &\ra& h/H/A + \chi^0_j
\label{neut-decay} \\
\chi^\pm_i &\ra& h/H/A + \chi^\pm_j \; .
\label{char-decay}
\end{eqnarray}
It is instructive to review how these decays
come about, and why the decay rate to Higgs bosons can be sizeable
throughout the kinematically allowed parameter space.

The centrally important gaugino-Higgs interactions are the 
kinetic terms of the Higgs supermultiplets.  They lead to
the component interaction terms~\cite{Martin:1997ns}
\begin{eqnarray}
& & -D_\mu H_u^\dagger D^\mu H_u^\dagger -
i \bar{\tilde{H}}_u \slashchar{D} \tilde{H}_u \nonumber \\
& & - \sqrt{2} g'\, Y_{H_u} \tilde{B} \tilde{H}_u H^*_u - \sqrt{2} g
\tilde{W}^a \tilde{H}_u t^a H^*_u + (u \leftrightarrow d) \; ,
\nonumber \\ 
\end{eqnarray}
where $Y_{H_u}$ is the hypercharge of the Higgs field.
The first term leads to ordinary gauge boson interactions
with the Higgs scalars,
\begin{eqnarray}
-(g')^2 B_\mu B^\mu Y^2_{H_u}\, H_u^\dagger H_u -
g^2 W^a_\mu W^{b\mu} \, {\rm tr} \left( t^a H_u^\dagger t^b H_u \right) 
\nonumber \\
+ (u \leftrightarrow d) \; . 
\end{eqnarray}
The second term leads to ordinary gauge boson interactions
with the Higgsinos,
\begin{eqnarray}
g'\, Y_{H_u} B_\mu \bar{\tilde{H}}_u \sigma^\mu \tilde{H}_u 
+ g\, W_\mu^a \bar{\tilde{H}}_u t^a \sigma^\mu \tilde{H}_u 
+ (u \leftrightarrow d) \; .
\end{eqnarray}
The latter are exactly the same interactions that Standard Model 
quark or lepton doublets have with Standard Model gauge bosons. 
Notice that the interaction involves the entire¶§
SU(2) doublets $H_u$,$H_d$, and thus, all eight real scalars 
\begin{eqnarray}
\mathrm{Re}(H_u^0), \mathrm{Im}(H_u^0), H_u^\pm, 
\mathrm{Re}(H_d^0), \mathrm{Im}(H_d^0), H_d^\pm \; .
\end{eqnarray}
In practice, as is very well known, linear combinations of 
the above become the physical mass eigenstate Higgs bosons
$(h,H,A,H^\pm)$ as well as the Goldstone bosons $(w^\pm,z)$ 
associated with the Standard Model gauge bosons $W^\pm,Z$.  
The critical observation is that the ``supersymmetrized'' 
interactions,
\begin{eqnarray}
- \sqrt{2} g' \tilde{B} \tilde{H}_u H^*_u - \sqrt{2} g \tilde{W}^a
\tilde{H}_u t^a H^*_u + (u \leftrightarrow d) \; , 
\label{critical-eq}
\end{eqnarray}
necessarily have gauge coupling strength to all components of
both Higgs doublets.  This means that, all other things considered 
equal, decays of $\tilde{B}/\tilde{W} \ra \tilde{H} + h/A/H$
lead to equal branching ratios into the different components 
of the doublet.  
A Higgs scalar, therefore, is just as common as a $W$ or $Z$
in this type of cascade decay.

In practice, the gaugino interaction eigenstates
mix with the Higgsino interaction eigenstates through the same
interactions, Eq.~(\ref{critical-eq}), that led to large decay
rates of gauginos into physical Higgs bosons.  
In the mass basis, the decays in Eqs.~(\ref{neut-decay}),(\ref{char-decay}) 
roughly translate into heavier neutralinos $\chi^0_{3,4}$ 
and charginos $\chi^\pm_2$ decaying into their lighter 
counterparts $\chi^0_{1,2}$, $\chi^\pm_1$ and Higgs bosons.  
The relevant branching ratios are thus  
$\chi^0_{4,3} \ra h/H/A + \left( \chi^0_1 \; \text{or} \; \chi^0_2 \right)$ and
$\chi_2^\pm \ra h/H/A + \chi_1^\pm$.
The other possible decay $\chi_2 \rightarrow h + \chi_1$ is mostly 
kinematically forbidden in the region of our interest. 

Another type of cascade decay occurs when the Higgsinos are
heavier than the winos and/or bino.  This opens up the decay
channels $\tilde{H} \ra  h/H/A + \tilde{B}/\tilde{W}$.
This might well provide an interesting source of Higgs bosons
given a cascade from third generation squarks to $\tilde{H}$.
Our preliminary work on this cascade suggests it takes more
luminosity than $10$~fb$^{-1}$ (which is the main focus of this
paper), and requires adjustments to the cut-based search
strategy to optimize for a signal of third generation squarks.

There are yet other superpartner cascade decays that could also 
lead to Higgs bosons, such as stop decay
$\tilde{t}_2 \ra \tilde{t}_1 + h/A/H$~\cite{  
Djouadi:1997xx, Djouadi:1999dg}.
To the extent that this process occurs for the specific points
in the MSSM parameter space we present below, it is included
in our inclusive analysis.  In practice, however, the production 
cross section of just the heavier stop $\tilde{t}_2$ is small relative to the
large number of other squarks (and gluino), 
while the branching ratio 
$\tilde{t}_2 \ra \tilde{t}_1 h$ is also accidentally small 
when $m_{\tilde{Q}_3} = m_{\tilde{t}_R}$ which we take for 
our Study Points.  This implies negligible numbers of lightest Higgs bosons
arise from $\tilde{t}_2$ decay.

Finally, there is an interesting possibility of cascade decays
into the charged Higgs $H^\pm$.  If $m_{H^\pm} > m_t + m_b$, 
the hadronic decay mode $H^+ \ra t \bar{b}$ often has a large fraction, 
and thus could be an interesting candidate for jet substructure 
techniques, utilizing top tagging
\cite{Kaplan:2008ie,agashe,gerbush,Brooijmans:2008zz,Thaler:2008ju,Almeida:2008yp,Almeida:2008tp,Ellis:2009su,Ellis:2009me,Krohn:2009zg,Plehn:2009rk}  
or other novel methods.

\section{Cascading to Boosted Higgs Bosons}
\label{sec:cascade}

\begin{figure}[t]
 \centering
 \includegraphics[width=0.45\textwidth]{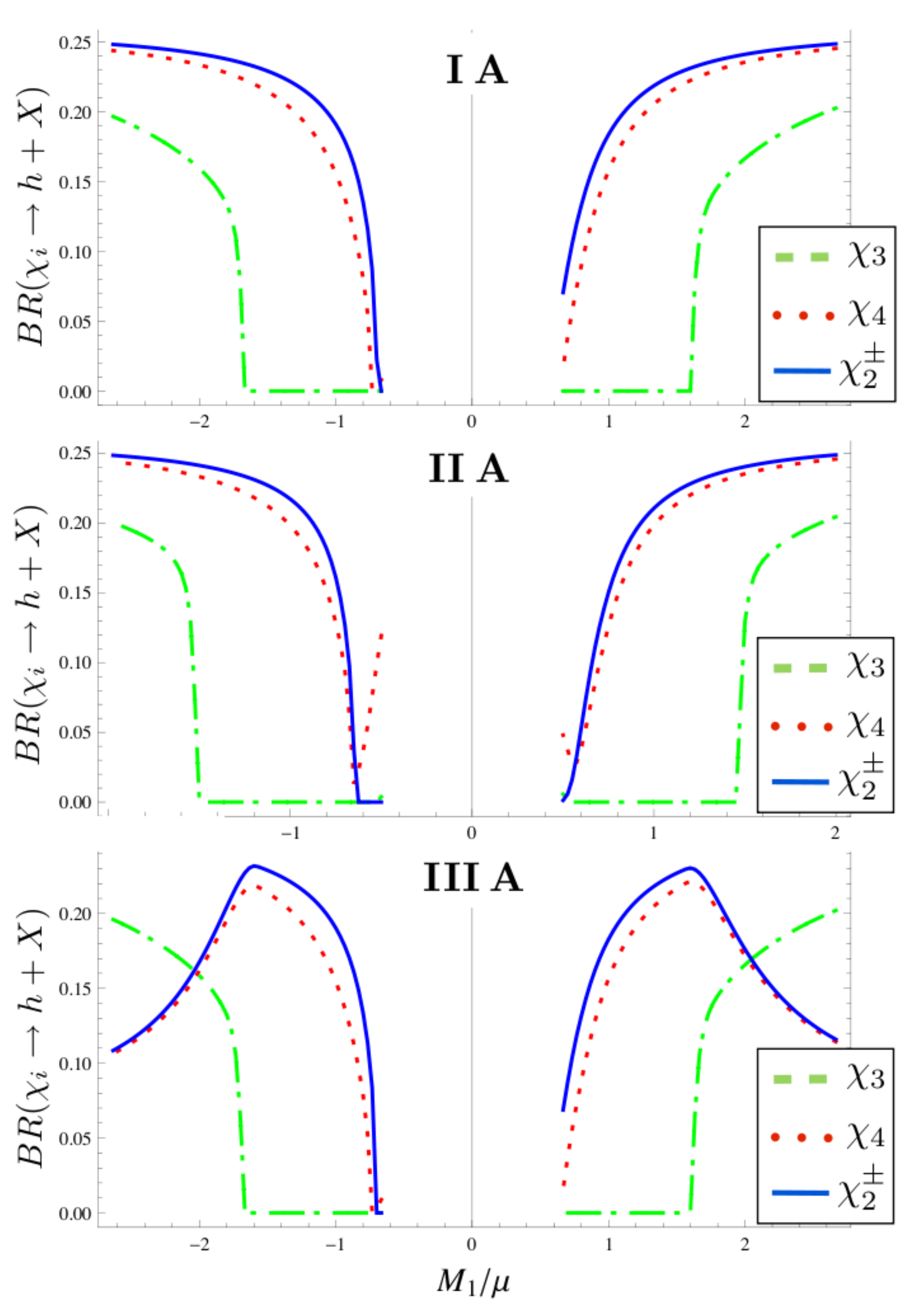}
 \caption{The branching ratios of heavier gaugino-like neutralinos 
and charginos into lighter Higgsino-like ones plus the lightest Higgs
boson is shown for the following parameters:  
We take $100 \; \mathrm{GeV} \; < M_1 = M_2/2 < 400$~GeV 
for all Figures, $\left|\mu\right| = 150~\gev$ 
in plots I and III and $\left|\mu\right| = 200~\gev$ in plots II. 
Plots I and II have heavier sleptons, $m_{\tilde{l}} > 800$~GeV,
so that two-body decays are kinematically forbidden.
In plot III, we take $m_{\tilde{l}} = 500$~GeV, 
which allows the wino to decay to left-handed sleptons
once $M_2 > 500$~GeV\@.
This is why the branching ratios of $\chi^0_4$, $\chi^\pm_2$ 
decrease above $M_1/|\mu| > 1.7$.}
\label{fig:br1a}
\end{figure}

\begin{figure}[!h]
 \centering
 \includegraphics[width=0.48\textwidth]{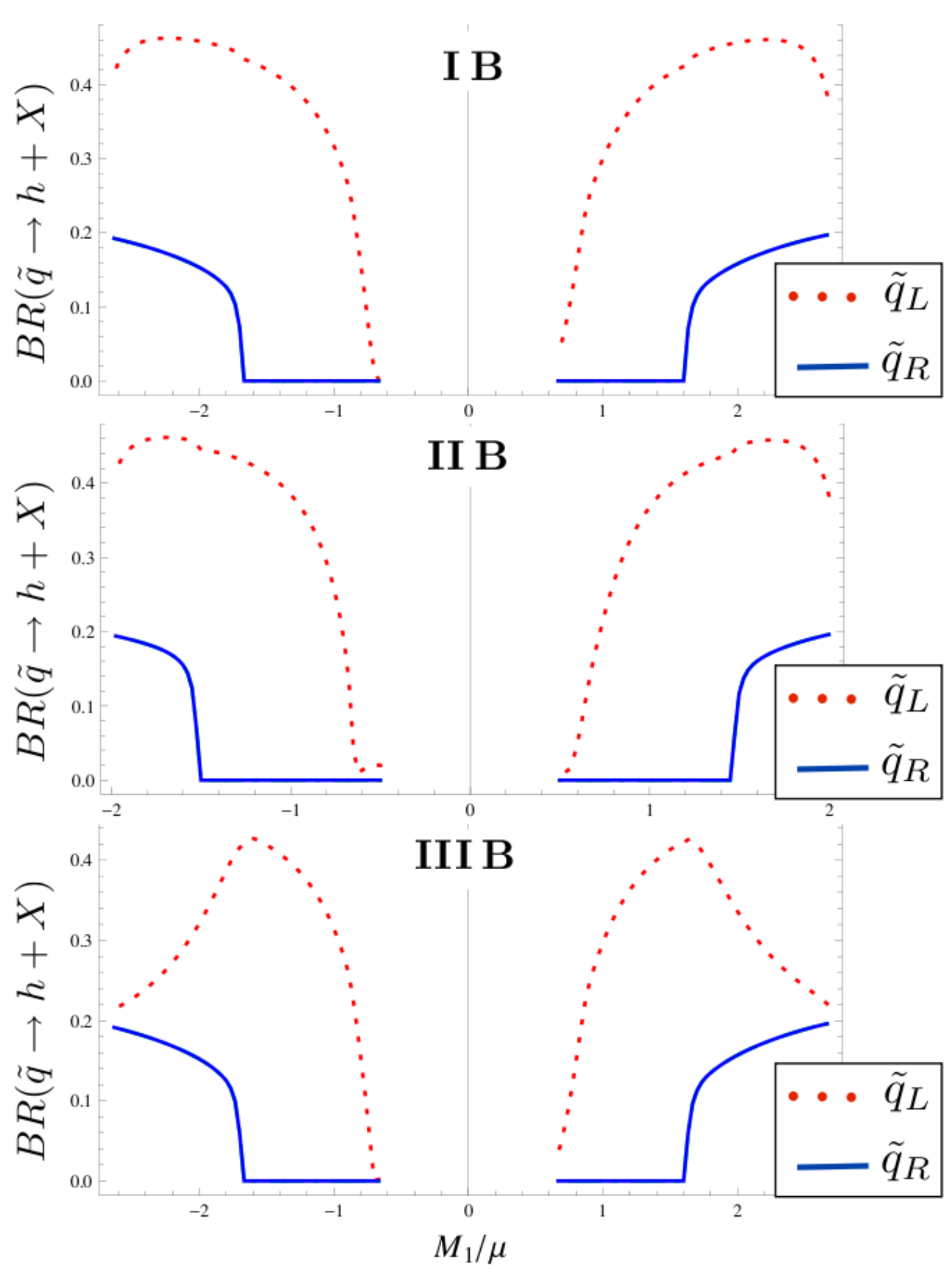}
 \caption{The branching ratios for decays to the lightest Higgs boson 
   as a function of $M_1/\mu$.  The MSSM parameters for each plot 
   are the same as the three rows in Fig.~\ref{fig:br1a}. Here $\tilde
   q_L$  refers to the sum of $\tilde u_L$ and $\tilde d_L$ (both
   components of the electroweak doublet), while $\tilde q_R$ refers
   to either $\tilde u_R$ or $\tilde d_R$. }  
\label{fig:br1b}
\end{figure}

The largest rate for Higgs boson production arises when 
first or second generation squarks cascade decay through gauginos, 
which then decay into lighter Higgsinos and Higgs bosons.  Generally,
first and second generation squarks decay as
\begin{eqnarray}
\tilde{q}_L &\rightarrow& q + \tilde{W} 
\label{sqL-eq} \\
\tilde{q}_R &\rightarrow& q + \tilde{B} \, 
\label{sqR-eq}
\end{eqnarray}
so long as the wino and bino satisfy the simple kinematical
requirement that they are lighter than the squarks.
The left-handed squarks can also decay to the bino,
but this rarely happens when the wino mode is kinematically open, 
since the ratio of bino to wino couplings for the left-handed
squark doublet is proportional to $Y_Q^2 (g'/g)^2 \simeq 0.01$.
Thus, to very good accuracy, first and second generation left-handed 
squarks decay through the wino, right-handed squarks decay 
through the bino.

Given a Higgsino plus Higgs boson lighter than the wino and/or bino, 
the two-body decays into Higgs bosons discussed in 
Sec.~\ref{sec:goldstone} become applicable.
Since the lighter quarks $(u,d,c,s)$ have Yukawa couplings 
far subdominant to the gauge coupling strengths $g,g'$, 
the cascade in which squarks decay directly into the Higgsinos
essentially never occurs.
This implies the large QCD-dominated production cross sections
of squarks can lead to substantial numbers of Higgs bosons
from the cascade decays with only modest mass hierarchy requirements.
Moreover, in addition to squark pair production, squark-gluino and 
gluino pair production can also lead to substantial rates of squarks 
through the two-body $\tilde{g} \ra \tilde{q} q$.

We have thus clearly demarcated the superpartner cascades into 
Higgs bosons as having two largely independent sources, namely
\begin{eqnarray}
\tilde{q}_L \ra q + \tilde{W}; \;\; \tilde{W} \ra h/H/A + \tilde{H} \\
\tilde{q}_R \ra q + \tilde{B}; \;\; \tilde{B} \ra h/H/A + \tilde{H} \; .
\end{eqnarray}
This will be useful as we consider variations of gaugino
masses and slepton masses.

Gauginos can also decay to sleptons, and it is fairly easy 
to see what effect they have if they are lighter than the
bino and/or wino.  For simplicity, consider all three generations
of sleptons to be degenerate.  If all sleptons are lighter than
the bino, we can estimate the branching ratio by just summing
over three generations of right-handed and left-handed leptons
plus one Higgs doublet.  We get
\begin{eqnarray}
BR(\tilde{B} \ra h \tilde{H}^0) & \simeq & \frac{1}{4} \, 
\frac{Y_H^2}{3 Y_L^2 + 3 Y_e^2 + Y_H^2} \; \simeq \; 0.015 \\
& &{} (\mathrm{for} \; m_{\tilde{l}_{L,R}} < M_1) \nonumber
\end{eqnarray}
where the $1/4$ comes from for picking just $h$ from $(h,z_0,w^+,w^-)$,
assuming $m_A$ is large.  (It is easy to generalize for smaller
$m_A$.)  If left-handed sleptons are lighter than the wino, 
we get
\begin{eqnarray}
BR(\tilde{W}^0 \ra h \tilde{H^0}) & \simeq & \frac{1}{4} \, 
\frac{Y_H^2}{3 Y_L^2 + Y_H^2} \; \simeq \; 0.06 \\
& &{} (\mathrm{for} \; m_{\tilde{l}_L} < M_2) \; . \nonumber
\end{eqnarray}
So, the effect of all sleptons lighter than the bino is to very
efficiently suppress the cascade decay of right-handed squarks
to Higgs bosons from about $1/4$ to about a percent.  
Conversely, the effect of left-handed sleptons lighter than 
the wino is to reduce the branching ratio from about $1/4$ to $1/16$.  
While this suppression is significant, it certainly does not 
eliminate this decay mode, and illustrates the robustness of
finding a Higgs boson within a fairly generic superpartner cascade.

We can study the branching ratios in more detail numerically.
As we have already seen, 
the likelihood of finding a Higgs boson in a complex decay chain 
originating from a squark can be approximated, to a large extent, 
by the product: 
\begin{eqnarray}
  \label{eq:decay-sq}
  P_{\tilde q h} \ &\equiv&  \ 
\text{Br} \left( \tilde q \rightarrow  X  \rightarrow h + Y \right)
\  \approx \\ 
& &  \sum_{\chi_a = \chi^0_{4,3}, \chi_{2}^\pm}  
   \text{Br}  \left( \tilde q \rightarrow \chi_a + \dots \right)  
   \text{Br}  \left( \chi_a \rightarrow h + \dots \right) \;, \nonumber
\end{eqnarray}
where $X$ and $Y$ are other particles or superpartners.

Consider now two interesting regimes for the masses of the
Higgs bosons.  The first, ``large $m_A$'', and the second, ``smaller $m_A$''.  

\subsection{Large $m_A$}

The first regime we consider is when
\begin{eqnarray}
m_A &\gg& \mathrm{min}(m_h,|M_2 - \mu|,|M_1 - \mu|) \; , 
\end{eqnarray}
often described as the ``decoupling limit''.
In this limit, all of the Higgs bosons $H$,$A$,$H^\pm$ are
predominantly eigenstates from the second Higgs doublet.
These Higgs bosons are much heavier than the lightest Higgs
as well as the lighter superpartners in the model.
In practice, we take $m_A \sim 1~\tev$, and thus 
the scalars $H, A, H^\pm$ have masses $\sim 1~\tev$, while 
$h$ mixes minimally to $H$ with ordinary couplings to 
standard model particles.

\subsubsection{Higgs in a cascade:}

In Fig.~\ref{fig:br1a} we show the branching ratios of 
$\chi^0_{4,3} \ra h + \chi^0_{1,2}$,
$\chi_2^\pm \rightarrow h + \chi_1^\pm$ and in Fig.~\ref{fig:br1b} we show 
$P_{\tilde q_L h}$ and  $P_{\tilde q_R h}$ 
as a function of $M_1/\mu$ in the large $m_A$ regime.

Plot I in Figs.~\ref{fig:br1a},\ref{fig:br1b} 
is generated with all the squarks and sleptons set to 
$1$~TeV\@.  Because of using a small value of $|\mu|$  (namely,
$150$~GeV), winos are relatively heavier than the Higgsinos and
mix minimally to the rest of the gauginos throughout. 
As a result, the heavier mass 
eigenstates ($\chi^0_4$ and $\chi^\pm_2$) are mostly winos. As
indicated in Plot I in Fig.~\ref{fig:br1a}, winos decay significantly 
to the lightest Higgs boson.  In fact,
for $M_2 \gtrsim 300~\gev$, wino decay follows the
``Goldstone region'':  roughly $3/4$ of the time the wino decays into
longitudinal $W/Z$ and $1/4$ of the time it decays into the 
lightest Higgs boson.  For a large part of the parameter
space the mass gaps between $\chi^0_3$ and $\chi^0_{1,2}$ are not large
enough to allow a two-body decay into the lightest Higgs boson. 
Once outside the kinematically forbidden zone, however, 
the branching ratio of the decay $\chi^0_3 \rightarrow h + \chi^0_{1,2}$ 
rises quickly with increasing $M_1$. In this region, $M_1/\mu$ 
is large and $\chi^0_3$ is mostly a bino.

The same spectrum is used to generate plot I in Fig.~\ref{fig:br1b}.
Note that the right-handed squarks decay mostly to the bino and 
so $P_{\tilde q_R h}$ looks almost identical to the branching ratio 
of $\chi^0_3$ decaying to Higgs boson. Similarly, the left-handed squarks decay 
mainly to the winos and $P_{\tilde q_L h}$ follows the partial decay 
width of $\chi^0_4$ and $\chi^\pm_2$ to Higgs bosons.  The other feature 
to note in this plot is that $P_{\tilde q_L h}$ goes down for 
large $M_1$, signifying that the decays of squark to quark plus wino 
are beginning to be affected by kinematical suppression from the
heavy wino.

Plots II in Figs.~\ref{fig:br1a},\ref{fig:br1b} 
are similar to Plots I except that slightly heavier
Higgsinos ($\left| \mu \right| = 200~\gev)$ are used.  Larger $M_1/\mu$
is needed in order to open up $\chi^0_3$ decays to Higgs bosons. 
A curious rise is seen in $\chi^0_4$ decays for small $M_1$. 
It is an artifact of decays $\chi^0_4 \rightarrow W^\pm + \chi^\mp_1$
shutting down, thereby, causing total decay width of $\chi^0_4$ to shrink. 
This feature is more prominent for negative $\mu$. Even if all
parameters in the chargino mass matrices are held fixed, taking
$M_2$ to have the opposite sign of $\mu$ reduces the splitting 
among the mass eigenvalues. 
This results in heavier $\chi^\pm_1$ and prevents the
two-body decay $\chi^0_4 \rightarrow W^\pm + \chi^\mp_1$ decays 
even for heavier $\chi^0_4$.  
Once again,  $P_{\tilde q_L h}$ and $P_{\tilde q_R h}$
follows the partial decay width of winos and bino respectively. One
thing to note is that even though there is a sharp rise in $\chi^0_4$
and $\chi^\pm_2$ partial widths for small $M_1$, there is no such
curious feature in $P_{\tilde q_L h}$. In this limit,  $M_2 \sim
\left| \mu \right|$ and decays of squarks to $\chi^0_4$
and $\chi^\pm_2$ suffer because of rising Higgsino content in them. 

Finally, for Plot III in Figs.~\ref{fig:br1a},\ref{fig:br1b}, 
all parameters are the same as Plots I except that the 
sleptons are taken be lighter, in this case, $500$GeV\@. 
As the wino mass is increased above this value, the
wino-like neutralino and charginos begin to decay into
slepton modes, reducing the branching fraction to the
lightest Higgs boson.

\subsubsection{Boost of a Higgs boson in a Superpartner Cascade} 

\begin{figure}[!h]
 \centering
 \includegraphics[width= 0.5\textwidth]{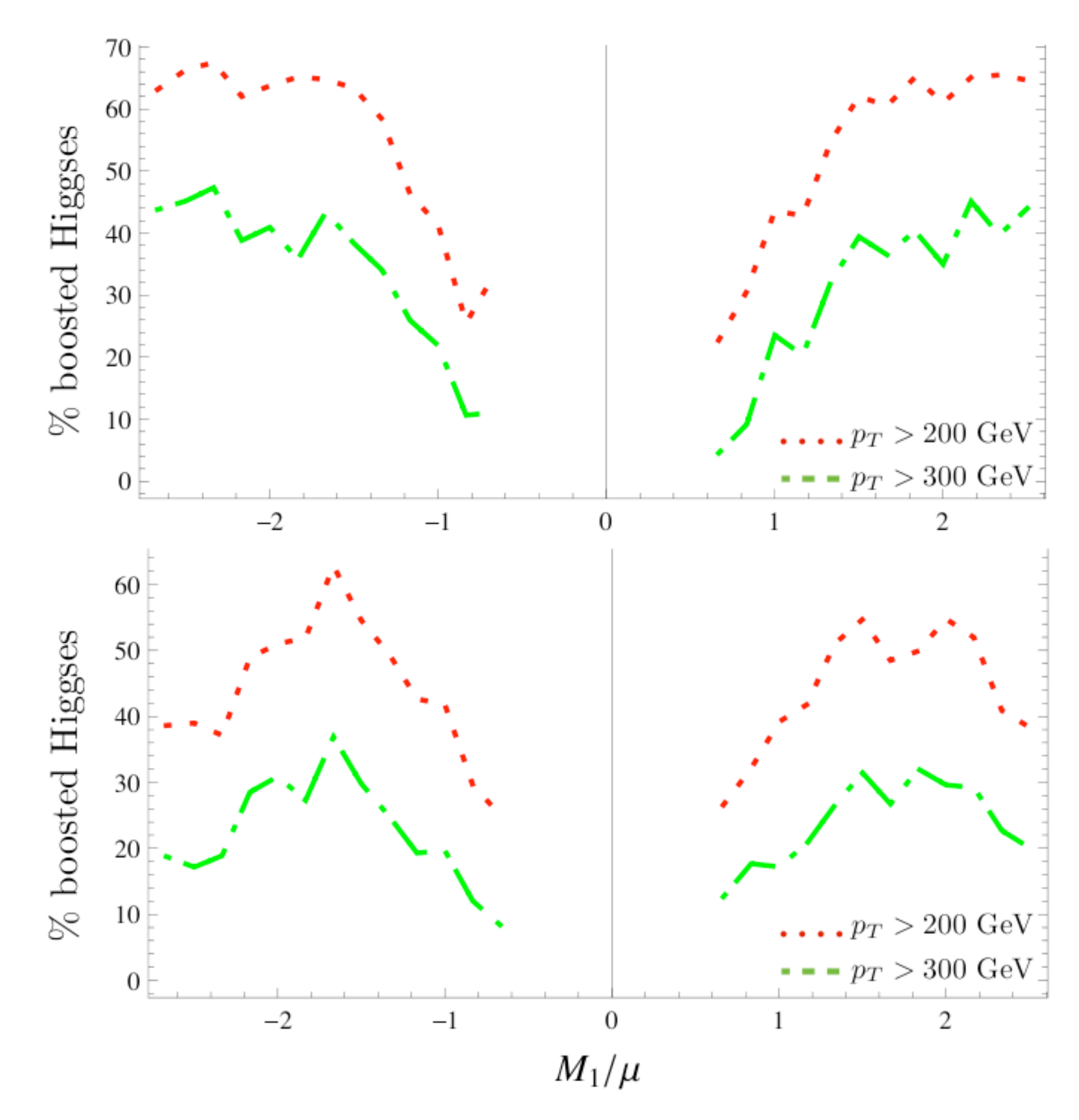}
\vspace{-0.9cm}
 \caption{The fraction (in \%) of boosted Higgs bosons as a function of
 $M_1/\mu$ with $M_2 = 2 M_1$, $\mu = 150~\gev$ and $\tan\beta = 10$ 
 in samples of events generated by PYTHIA.
 In the plots the red and dotted lines represent the percentages of
 Higgs bosons with $p_T > 200~\gev$ and the green dot-dashed lines
 represent the fraction of Higgs with $p_T > 300~\gev$. In the left
 Figure the squark masses are $1~\tev$, while in the right Figure the
 squark masses are $750$~GeV\@.  All other relevant soft supersymmetric
 breaking masses are kept at or above $1$~TeV\@.} 
\label{fig:boost}
 \end{figure}

In a typical cascade, a Higgs boson appears from the decay of a massive
superpartner.  The large release in rest mass results in a large
recoil energy, i.e., Higgs bosons from superpartner decays are
naturally boosted.  This is demonstrated in Fig.~\ref{fig:boost},
which shows that a significant fraction of Higgs bosons are boosted 
with $p_T > 200~\gev$ (and even with $300~\gev$, as shown). 
The boost was found by generating samples of $5000$ supersymmetric events 
at different values of $M_1/\mu$ using PYTHIA~v6.4~\cite{Sjostrand:2006za},
and plotting the Higgs transverse momenta.

Both of the plots in Fig.~\ref{fig:boost} are made with 
$\mu = 150~\gev$, $\tan\beta = 10$ and all squarks with mass of 
$1~\tev$.  Sleptons have mass of $1~\tev$ in plot I and 
$500~\gev$ in plot II\@. The presence of light sleptons reduces the
fraction of supersymmetric production that leads to a boosted Higgs boson 
in the cascade.  This is due not only does the overall lower 
fraction of Higgs bosons appearing in the cascades 
(see plot IIIA and IIIB in Fig.~\ref{fig:br1b}) when heavy neutralinos 
and charginos decay to them, but also fewer of the Higgs bosons 
in the decay chain are boosted.  

\subsection{Smaller $m_A$}

The second interesting regime of the Higgs sector that we consider
is smaller $m_A$, where 
\begin{eqnarray}
m_A &<& \mathrm{min}(|M_2 - \mu|,|M_1 - \mu|) \; . 
\end{eqnarray}
There are really two distinct regimes of smaller $m_A$:
the first is when all the Higgs mass eigenstates 
($h, H, A \text{ and } H^\pm$) are comparable in mass and the 
CP even neutral Higgs scalars $h$ and $H$ mix maximally among 
each other. 
The second is when there is less mixing, but $H, A, H^\pm$ 
are light enough to be kinematically accessible to gaugino decay. 
We will examine both of these cases below.

Interestingly, the branching ratios for $H \ra b\bar{b}$
and $A \ra b\bar{b}$ remain the dominant channels decay modes
for modest (or larger) $\tan\beta$ even when decays
to gauge boson pairs becomes kinematically accessible. 
For $m_A \gg m_Z$, the mixing angle $\tan 2\alpha \ra \tan 2\beta$,
and thus $H$ is mostly $H_d^0$.  Larger $\tan\beta$ implies
$\langle H_d^0 \rangle \ll \langle H_u^0 \rangle$, 
and thus the 3-point couplings $H W_\mu W^\mu$, $H Z_\mu Z^\mu$
are suppressed.  Analogously, since there is no expectation
value for the CP-odd scalar, these 3-point couplings
are exactly zero.  Thus, the decays into $b\bar{b}$ remain dominant
until $m_{H,A} \lsim \mathrm{min}(2 \mu,2 M_1,2 M_2)$, 
where decays into the lightest gauginos becomes 
kinematically accessible.  This suggests that the 
$H,A \ra b\bar{b}$ mode is viable up to well past
$200$~GeV (twice the smallest allowed Higgsino mass), 
and for the Study Points in this paper, up to and
beyond $300$~GeV\@. 

\begin{figure}[!h]
\vspace{- 0.1cm}
 \centering
 \includegraphics[width= 0.45\textwidth]{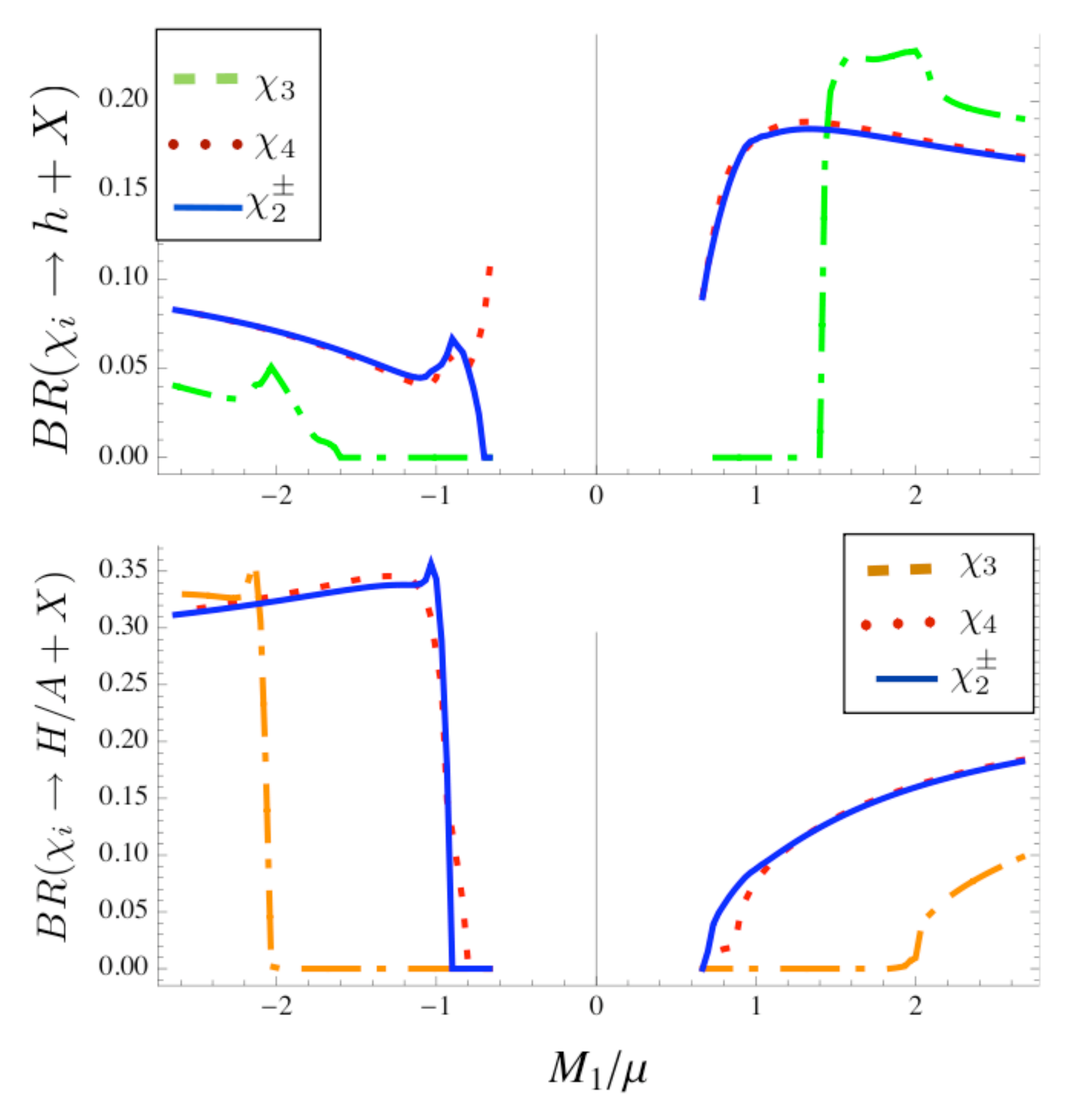}
 \caption{The branching ratio for decays to a Higgs boson 
   is shown as a function of
   $M_1/\mu$ for $m_A = 150~\gev$, $\left|\mu\right| = 150~\gev$, 
   and $\tan\beta = 4$. 
   The upper plot shows the decay rates of heavy gauginos into the
   lightest Higgs boson, 
   while the lower plot shows the summed decay rates to the heavier 
   Higgs bosons $H/A$. The squark and slepton
   masses are taken to be $1$~TeV\@.} 
\label{fig:br2a}
\end{figure}

\begin{figure}[!h]
\vspace{- 0.1cm}
 \centering
 \includegraphics[width= 0.45\textwidth]{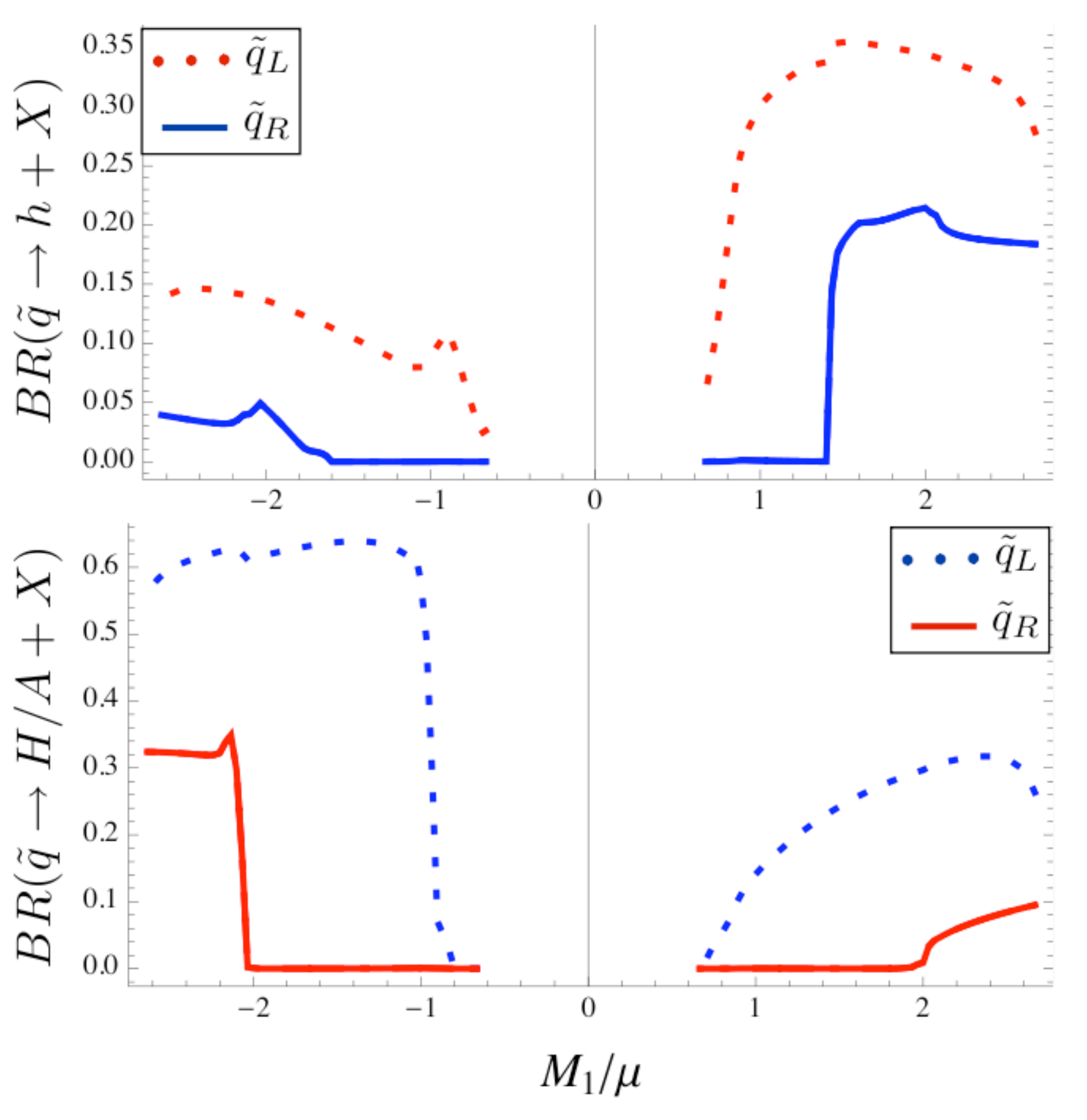}
 \caption{The branching ratio for squark decays to a Higgs boson 
   as a function of
   $M_1/\mu$ for $m_A = 150~\gev$, $\left|\mu\right| = 150~\gev$, 
   and $\tan\beta = 4$. 
   The upper plot shows the decay rate to the lightest Higgs boson,
   while the lower plot shows the summed decay rate to the heavier 
   Higgs bosons $H/A$. As in Figure.~\ref{fig:br1b},  $\tilde q_L$
   refers to the sum of $\tilde u_L$ and $\tilde d_L$, while $\tilde
   q_R$ refers to either $\tilde u_R$ or $\tilde d_R$.The squark and
   slepton masses are taken to be $1$~TeV\@.} 
\label{fig:br2b}
\end{figure}

\section{Mixed Higgsino/bino as Dark Matter}
\label{sec:relic}

\begin{figure}[!h]
 \centering
 \includegraphics[width= 0.45\textwidth]{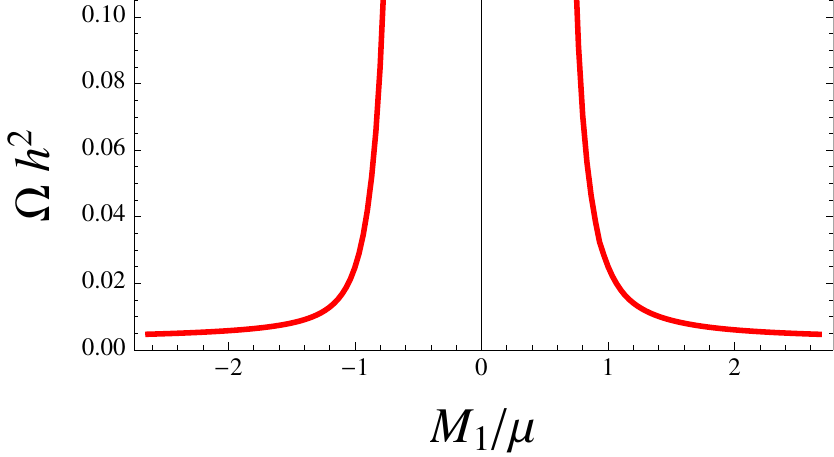}
 \caption{ The LSP relic density as a function of $M_1/\mu$ with
   $M_2 = 2 M_1$, $|\mu| = 150$~GeV, and $\tan \beta = 10$. 
   The squarks, sleptons, and $m_A$ were taken to be $1$~TeV\@.
   The thermal relic density was calculated using 
   micrOMEGAs~v2.4 \cite{Belanger:2010pz}.}
 \label{fig:relic}
 \end{figure}

One of the more attractive features of the weak
scale supersymmetry with conserved $R$-parity is that there exists 
a stable, neutral, colorless, weakly-interacting particle near the
electroweak scale.  In the post-LEP era, however, the prediction of 
present dark matter density does not automatically agree with the 
observation.  LEP bounds typically forces the superpartner spectrum 
to be heavier and hierarchical.  In this scenario,
neutralinos are closer to pure gauge eigenstates, namely bino, wino
and Higgsinos.  Avoiding coannihilation and Higgs pole regions, 
the relic density is generally too large for a bino and too small for
Higgsinos and winos.  Matching cosmological data seemingly requires 
rather precise relations among supersymmetry breaking parameters
(see e.g.\ \cite{ArkaniHamed:2006mb}).

Much of these constraints follow from the requirement  
that the LSP relic density matches the observed cosmological
dark matter density.  For our purposes, we are content to simply not
predict too much dark matter, since other non-thermal sources of
dark matter may be present.  All of the Study Points considered
in the paper automatically have a thermal LSP relic abundance that 
is at or smaller than the observed cosmological abundance,
$\Omega_{\chi^0_1} h^2 \leq 0.1$.  

In Fig.~\ref{fig:relic} we show the calculated thermal relic density 
$\Omega_{\chi^0_1} h^2$ is plotted as a function of $M_1/\mu$ 
for a fixed value of $|\mu| = 150~\gev$ and $\tan\beta = 10$. 
The thermal relic density was calculated using 
micrOMEGAs~v2.4 \cite{Belanger:2010pz}.
All squarks and sleptons were taken to be $1~\tev$. 
This clearly shows the variation of $\chi^0_1$ relic density 
as the gaugino/Higgsino content of LSP is changed\@.  
For large values of $M_1/\mu$, the lightest neutralino
is mostly a Higgsino.  As is well known, Higgsino-like neutralinos 
annihilate efficiently into gauge bosons, causing the calculated 
relic density to be smaller than the cosmological density. 
As the bino fraction of $\chi^0_1$ increases with decreasing $M_1$, 
the annihilation rate goes down, and thus relic density goes up.  
Since the squarks and sleptons are much heavier than the gauginos, 
the bino rarely annihilates through them.  
For the specific parameters we considered, we find
the annihilation rate can be optimized to give the right relic
abundance to match the observed cosmological abundance
when $M_1 \sim \left| \mu \right|$.

Matching the thermal relic density by taking $M_1 \sim |\mu|$ means bino 
cannot decay into Higgsinos and Higgs bosons.
Given the near independence of $\tilde{q}_L \ra \tilde{W}$ and 
$\tilde{q}_R \ra \tilde{B}$ (c.f.\ Sec.\ref{sec:cascade}), only roughly
half of the potential Higgs signal is lost given that right-handed
squarks no longer lead to decays into Higgs bosons.  
We present one Study Points that demonstrates the Higgs signal 
remains perfectly viable when $M_1 \sim |\mu|$.

\section{Jet Substructure Algorithm}
\label{sec:substructure} 

There are now several interesting techniques that exploit 
jet substructure to enable better identification of 
standard model or beyond-the-SM signals~\cite{Butterworth:2007ke,
  Brooijmans:2008zz, Butterworth:2008iy,Kaplan:2008ie,
  Almeida:2008tp,Thaler:2008ju,
  Butterworth:2009qa,Plehn:2009rk,Kribs:2009yh,
  Ellis:2009su,Ellis:2009me,Krohn:2009th,Soper:2010xk}. 
The central idea motivating the elaborate jet manipulation is that it is
possible to seek a single ``fat jet'' (that is, a jet with a particular
structure consistent with one coming from a massive particle decay) from 
the decay products of a boosted particle.
Butterworth, Davison, Rubin, and Salam (BDRS) \cite{Butterworth:2008iy} 
demonstrated that the Higgs boson of the Standard Model 
could be found with high significance using this 
technique \cite{Butterworth:2008iy}.  Their particular study
has been validated by a realistic simulation done by the 
ATLAS collaboration \cite{atlas-study-vh}.

The substructure algorithm developed by BDRS \cite{Butterworth:2008iy} 
to find a Higgs boson has two distinct parts:  
First determine whether a jet contains substructure consistent with 
coming from a Higgs decay to $b\bar{b}$. 
If it passes the criteria,
``filter'' the jet, improving the resolution of the invariant mass 
of the candidate resonance jet significantly.  
In order to identify a jet as a ``fat jet'', BDRS
stipulate two conditions:
the mass of individual subjets are significantly smaller than the mass
of the jet (the mass-drop condition) and the splitting of the jet into
the two subsets is not too asymmetric. The mass-drop condition
basically checks how the jet-mass is distributed in the jet-area, and
seeks out a jet that is consistent with one accommodating all the decay
products of a massive particle. Given the immense rate for QCD jets, 
the mass-drop condition alone is not enough. The background jets are, 
however, dominated by gluon splittings which exhibit soft and 
collinear singularities. These singularities imply the majority of 
QCD subjets are asymmetric, so by rejecting particularly 
asymmetric splittings, the background can be further suppressed. 

\subsection*{Substructure for Supersymmetry}

In Ref.~\cite{Kribs:2009yh} we proposed an algorithm to extract 
a Higgs boson signal using its dominant decay mode, $h \ra b\bar{b}$ 
from a new physics event sample. Our algorithm 
exploits the techniques developed by BDRS, with some additional
steps designed to allow our algorithm to be somewhat more efficient
than BDRS when applied to busier final states characteristic of 
new physics. Following the criteria laid out in
Sec.~\ref{sec:cascade},  the simplest superpartner cascade which
yields a boosted Higgs is,  
$\tilde q \rightarrow \chi + j \rightarrow \chi' + h + j$, which
necessarily involves  
one additional hard parton. More complicated signal events, with
multiple extraneous, hard partons are easy to imagine. 
These hard partons, and their associated showers, can end up in the 
same fat-jet as the $h \ra b\bar{b}$. 
As these contaminating partons come from heavy particle decay 
and not from QCD radiation, they can survive the mass-drop 
and the asymmetry cuts (top and bottom quarks coming from the decay 
of superpartners are particularly dangerous as they also possess 
heavy flavor). Consequently, while declustering a fat jet, 
one may encounter multiple stages (say, ``thresholds'') 
that would pass all substructure criteria cited above. 

The BDRS algorithm is designed to consider only the \emph{first} 
declustering stage that satisfies the mass-drop and asymmetry conditions,
and as such, it is more susceptible to false thresholds encountered
in new physics events. 
BDRS jets are built using the inclusive C/A algorithm
\cite{Dokshitzer:1997in,Wobisch:1998wt,Wobisch:2000dk}, where subjets  
closer to each other are combined earlier, so the first threshold 
encountered will be where the subjets are maximally separated. 
To help distinguish between real and false thresholds, we need 
to use more information about the subjets. 
Although the contaminating hard partons are not removed by the mass-drop 
and asymmetry cuts, they necessarily introduce a new scale 
into the jet. 
Rather than select a threshold based on separation alone, we select 
the threshold where the subjet kinematics are maximally similar. 
Specifically, we impose a measure of similarity: maximize 
subjet hardness weighted by the inter-subjet separation. 
This measure takes advantage of the isotropic decay scalar
particles -- the Higgs bosons -- that we are interested in. 
The algorithm \cite{Kribs:2009yh} is described in full detail 
below.

\subsection{Our Algorithm}

The first step in our algorithm is to group final state particles,
after all showering and hadronization, into ``cells''  of size $\Delta
\eta \times \Delta \phi = 0.1 \times 0.1$\@. All particles in a cell
are combined, and the three-momentum of the total is rescaled such
that each cell has zero invariant mass~\cite{Thaler:2008ju}. Cells
with energy $< 1\ \gev$ are discarded, while the rest are clustered
into jets. The initial clustering is done using the inclusive C/A
algorithm, as implemented in FastJet~\cite{Cacciari:2005hq}, and
taking the jet size to be $R = 1.2$. Once the jets are formed, we
search for heavy flavor; this is an essential step given that we want
to discover the Higgs through its decays to bottom quarks. We b-tag
jets by looking through the event record for b-mesons or b-baryons. If
there is a b-flavored object within $20^{\circ}$ of a jet direction,
we tag the nearby jet as a b-jet with $60\%$ probability. If there are
no b-flavored objects in the vicinity of the jet in question, the jet
is tagged as a b-jet with a ``fake-rate" of $2\%$. Every b-tagged jet
in the event is then decomposed to search for substructure following
the steps below: \\

\noindent \textbf{1.}  Undo the last stage of jet-clustering. As a jet
is built from a sequence of $2\rightarrow 1$ mergings, unclustering
one stage yields two subjets. The two subjets  
$j_1$ and $j_2$ are labeled such that $m_{j_1} > m_{j_2}$\@. \\  
\textbf{2.} Following Ref.~\cite{Butterworth:2008iy}, 
subjets are checked for the existence of a significant mass drop 
($m_{j_1} < \mu\, m_j$), as well as non-existence of an asymmetry defined by
$y = \frac{\text{min}\left( p_{T_{j1}}^2,  p_{T_{j2}}^2 \right) }{m_j^2}
\Delta R^2_{j_1,j_2} > y_\text{cut}$.  We use 
$\mu = 0.68$ and $y_\text{cut} = (0.3)^2$ identical to 
Ref.~\cite{Butterworth:2008iy}.  
Both subjets are required to be $b$-tagged and have $p_T > 30\
\gev$\@. If these conditions are satisfied, this stage of clustering
(say, $i$-th)  
is recorded and then the following is calculated:
\begin{equation}
    \label{eq:discriminant}
    S_i = \frac{\text{min}\left( p_{T_{j_1}}^2,  p_{T_{j_2}}^2 \right) }
       {\left( p_{T_{j_1}} +  p_{T_{j_2}} \right)^2 } \Delta R_{j_1
         j_2} \; .
\end{equation}
The quantity $S_i$ (namely, similarity) is an indicator of the
similarity of the  
two subjets and is weighted by their separation $\Delta R_{j_1
  j_2}$. \\
\textbf{3.} Replace $j$ by $j_1$ and repeat from step~$1$ as long as $j$
has further subjets. \\
\textbf{4.} Select the stage of clustering for which $S_i$ is the
largest.  We anticipate that the two $b$-tagged subjets, at this stage,
are most likely to have originated from Higgs decay since they are 
more likely to be similar to each other.  If the two C/A $b$-tagged
subjets originate from Higgs decay, the subjets with opening 
angle $\Delta R_{j_1 j_2}$ should contain all the perturbative
radiation from the $b\bar{b}$ system by virtue of angular
ordering \cite{Bassetto:1984ik}.  However, the subjets still tend 
to include too much contamination from underlying events.  We then
filter \cite{Butterworth:2008iy} the events:
we cluster the jet constituents again using a finer angular scale 
specific to the jet [we use, $\text{min} \left( R_{bb}/2, 0.3\right)$]
and retain only the three hardest components $(b \bar{b} g)$.
Finally, we combine the three subjets and call the resultant 
a ``candidate resonance jet''. 

\subsection{Comparison with BDRS Algorithm}

Our algorithm declusters the event entirely, thereby checking 
multiple thresholds, while the BDRS algorithm only checks a 
single threshold. In an environment where there are few extraneous 
partons flying around, such as  $W^{\pm}H$ production or even 
supersymmetric Higgs production from short cascades, there are 
few false thresholds and the two algorithms perform comparably. 
However, as the number of extra partons (and thus the number of 
false thresholds) increases, there is a clear difference in efficiency. 
Any threshold, genuine or not, will stop the BDRS algorithm, 
while our approach takes in all thresholds and sorts them out 
using the $p_T$ similarity. Events with a true threshold masked 
by a false threshold at larger $R$ will be missed by BDRS, 
but captured by our approach. Of course, the similarity variable 
will not always select out a true threshold from among several, 
so accuracy is not necessarily increased. Fig.~\ref{fig:comp}, 
shown below, is a simple demonstration of how our algorithm is 
more efficient in a crowded environment.

\begin{figure}[t]
 \centering
 \includegraphics[width= 0.4\textwidth]{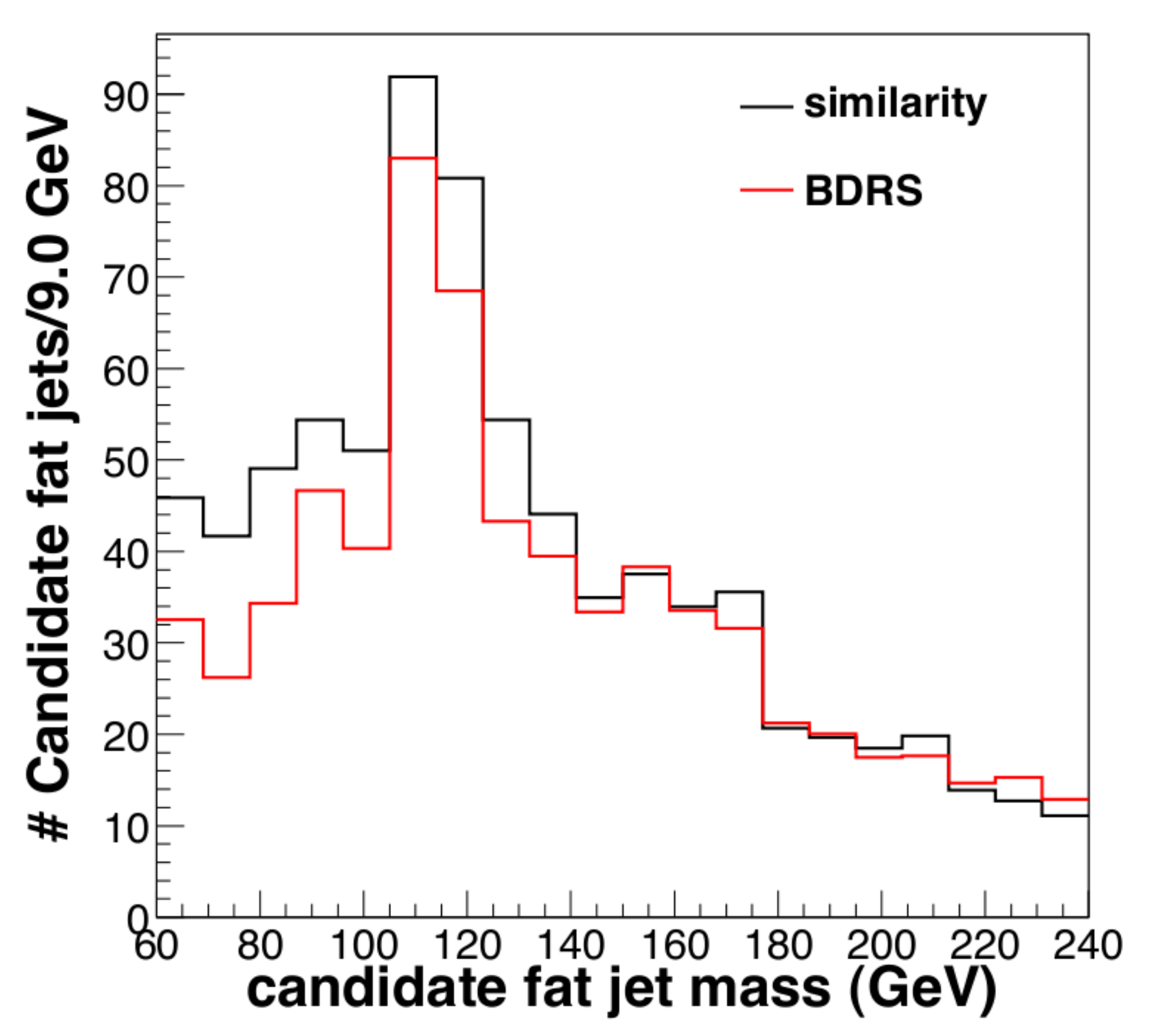}
 \caption{Comparison of the candidate resonance jet mass using the
   similarity algorithm (black) and the original BDRS algorithm (red),
   applied to our $b$-tagged fat jet sample. The signal point used for
   the comparison is SHSP 3 (see Table~\ref{table:points}), and the
   vertical axis has been rescaled to correspond to an integrated
   luminosity of $\mathcal L = 10\ \text{fb}^{-1}$. While the accuracy
   of the two algorithms is similar, the similarity algorithm is more
   efficient.}  
\label{fig:comp}
\end{figure}

This Figure was generated from a signal sample of $80$K
PYTHIA-generated events using the spectrum SHSP 3 listed in
Table~\ref{table:points}. The key feature of this spectrum is that the
gluino is as light as the squarks, so the signal sample has a
significant number squark-gluino associated-production events. The
gluinos decay through off-shell squarks, and typically lead to busy,
multi-jet events. As a measure of the increased efficiency, we can
count the number of event under the putative Higgs peak and compare
with the number of events in the bins adjacent to the peaks for each
of the algorithms. The significance, defined as (\# events in peak
above continuum)/$\sqrt{}$(\# continuum events) is larger for the
similarity algorithm by a factor of $\sim 1.2$. As we will later see,
the continuum supersymmetric events are often the largest background
to the Higgs peak, so we can expect the complete significance
(including SM backgrounds) to increase by roughly the same amount.

It is important to point out that, although we employ this algorithm
to find Higgs, all we  
really check for is a massive particle decaying to $2~b$ partons.
Any heavy multiplet that decays to $2~b$ should also be selected
by our jet algorithm as long as they are boosted. Among SM particles
we  expect to find $Z \rightarrow b\bar b$~\footnote{In practice, 
as light jets can occasionally fake $b$-jets, any boosted, 
heavy particle which decays hadronically ($t, W, \cdots$) 
has a chance of being picked up by the substructure algorithm.}. 
Also, when both $H$ and $A$ are light and decay to $b\bar{b}$, 
our algorithm can discover Higgs bosons as long as they are 
produced in a superpartner cascade.   

\section{Results}
\label{sec:results}

Having demonstrated sparticle cascade decays as a viable, important
source of boosted Higgs bosons and described our substructure
algorithm in detail, we now demonstrate the effectiveness of our
proposal. To best convey our results, we first propose a collection of
Study Points on which we use the candidate resonance jet finding
algorithm.  While by no means exhaustive, the Study Points have a
diverse set of MSSM  
parameters. After introducing the Study Points, we then
list the set of backgrounds we considered for this work and show the
way in which  
sets of conventional cuts can be used to reduce these. The
candidate resonance jet finding algorithm is then run on this set of
rarefied events (both signal and background events). Finally, masses
of the candidate resonance jets are plotted to estimate the signal
significance.

\subsection{Supersymmetric Higgs Study Points}
\label{subsect:shsp}

\noindent

The efficiency of our algorithm to find Higgs bosons is demonstrated on a set
of benchmark points, Supersymmetric Higgs Study Points (SHSPs), tabulated in
Table~\ref{table:points}. These Study Points are grouped into three categories.

\begin{itemize}
\item Study Points $1$,$2$ represent spectra in the decoupling limit
  ($m_A = 1~\tev$) . In SHSP~$1a$ and $1b$ the LSP is mostly bino, all 
squarks are at $1~\tev$, and the sleptons are at $1~\tev$ and
$350~\gev$ respectively. In SHSP~$2a$ and $2b$ the LSP is a maximal
mixture of Higgsinos and bino. In SHSP~$2a$ once again we use heavier
squarks and sleptons while  slightly lighter squarks and sleptons are used in 
SHSP~$2b$.  
\item SHSP 3 has $M_1 \simeq |\mu|$ and large $m_A$, such that the LSP
  has a thermal relic density that matches cosmological measurements.  
\item The final set of Study Points, SHSP~$4$,$5$ ($|\mu| = 150\
  \gev$) and SHSP~$6$ $(|\mu| = 200\ \gev)$ are 
representatives of spectra in the smaller $m_A$ region. The main
difference between SHSP~$4$,$6$ versus SHSP~$5$ is the sign of the
$\mu$ term.  As shown in Fig.~\ref{fig:br2b}, when $m_A$ is low the
sign of $\mu$ greatly influences which Higgs bosons the gauginos decay
into. For SHSP~$4,6$, decays to $h$ predominate, while $H/A$
predominate in SHSP~$5$. 
\end{itemize}

To simulate the supersymmetric signal, we use 
PYTHIA~v6.4 to generate parton 
level events, with subsequent showering and hadronization. The
lowest-order, inclusive superpartner production cross sections are
large ($\mathcal O(\text{pb})$) and are listed for all Study Points in
Table~\ref{table:points}.  These cross sections are somewhat
misleading, since the quoted cross sections also include electroweak
production of light charginos and neutralinos.  In the scenarios we
are considering, the lightest charginos and neutralinos have a large
Higgsino component and thus large couplings to the $Z$ boson. As a
result, the LHC cross sections for neutralino pair production
$\chi^0_1 \chi^0_2$, chargino pair production $\chi^{\pm}_1
\chi^{\mp}_1$ and associated chargino-neutralino production
$\chi^0_{1,2} \chi^{\pm}_1$ are all quite large, $\mathcal O(0.5-1
\text{pb})$. While a large chargino/neutralino production cross
section will likely enable the discovery of new physics, light
neutralinos and charginos do not decay to Higgs bosons so these events
are of no use for a Higgs search. Therefore, in order to fairly judge
our Higgs-finding algorithm, we have included the fraction of
supersymmetric events containing a Higgs boson ($h/H/A$) in
Table~\ref{table:points}. This fraction was calculated by counting the
number of on-shell Higgs bosons, without any kinematic cuts, in
samples of PYTHIA-generated supersymmetric events. The final row in
Table~\ref{table:points}, $\sigma_{h/H/A}$, is simply the inclusive
supersymmetric cross section times the fraction of supersymmetric
events containing a Higgs boson. 

\begin{centering}
\begin{table*}[!ht]
 \centering
 \begin{tabular}[c]{|c||c|c||c|| c | c|c|}
\hline
 & SHSP~1a / SHSP~1b & SHSP~2a / SHSP~2b 
      &  SHSP~3 &  SHSP~4 &SHSP~5 & SHSP~6 \\ \hline
 $\tan\beta$   & $10$       & $10$       
      & $10$   & $5$  & $6.5$  & 10 \\
 $M_1$         & $300~\gev$  & $150~\gev$  
      & $163~\gev$  & $200~\gev$ & $200~\gev$ & $300~\gev$ \\
 $M_2$         & $600~\gev$  & $300~\gev$  
      & $400~\gev$  & $400~\gev$ & $400~\gev$ & $600~\gev$\\
 $M_3$         & $2.1~\tev$  & $1.05~\tev$ 
      & $1.0~\tev$ & $1.4~\tev$  & $1.4~\tev$ & $2.1~\tev$  \\
 $\mu$         & $150~\gev$  & $150~\gev$  
      & $200~\gev$ &  $200~\gev$  & $-150~\gev$ & $150~\gev$\\
 $m_A$         & $1~\tev$    &   $1~\tev$  
      & $1~\tev$ & $150~\gev$  & $150~\gev$ & $200~\gev$  \\
 $a_t$         & $900~\gev$ & $-900~\gev$ 
      & $900~\gev$ &  $2.04~\tev$\footnote{$a_b = a_t$ 
       for this point as well} & $1.4~\tev$ & $900~\gev$ \\
 $m_{\tilde{q}}$ & $1~\tev$   & $1~\tev / 750~\gev$ 
      & $1~\tev$ & $1~\tev$ & $1~\tev $ & $1~\tev$ \\ 
 $m_{\tilde{l}}$ & $1~\tev / 350~\gev$  & $1~\tev / 350~\gev$
      & $350~\gev$ &  $1~\tev$ & $1~\tev$ & $1~\tev$\\ 
\hline
 $m_h$         & $116~\gev$  & $117~\gev$ 
      & $116~\gev$ & $114~\gev$ & $115~\gev$ & $115~\gev$\\
 $m_H$         & $1~\tev$  & $1~\tev$  
      & $1~\tev$  & $161~\gev$ & $157~\gev $ & $202~\gev$\\
 $m_A$         & $1~\tev$  & $1~\tev$  
      & $1~\tev$ &  $150~\gev$ & $150~\gev $ & $200~\gev$\\
 $m_{H^\pm}$    & $1~\tev$  & $1~\tev$  
      & $1~\tev$  & $169~\gev$ & $170~\gev$ & $216~\gev$\\
\hline
 $\chi_1$      & $138~\gev$  & $110~\gev$ 
      & $140~\gev$   & $157~\gev$ & $136~\gev$ & $138~\gev$\\
 $\chi_2$      & $-158~\gev$  & $-161~\gev$ 
      & $209~\gev$ & $-207~\gev$ & $-163~\gev$ & $-158~\gev$\\
 $\chi_3$      & $206~\gev$  & $174~\gev$ 
      & $-209~\gev$  & $227~\gev$ & $210~\gev$ & $306~\gev$\\
 $\chi_4$      & $625~\gev$  & $338~\gev$ 
      & $429~\gev$  & $433~\gev$ & $426~\gev$ & $623~\gev$\\
 $\chi_1^{+}$  & $148~\gev$  & $137~\gev$ 
      & $191~\gev$  & $187~\gev$ & $152~\gev$ & $148~\gev$\\
 $\chi_2^{+}$  & $625~\gev$  & $337~\gev$ 
      & $429~\gev$ &  $433~\gev$ & $426~\gev$ & $623~\gev$ \\
\hline
 $\sigma_{\text{tot}}$  & $3.9$~pb  & $5.8$~pb $/~ 8.07$~pb 
      &  $2.76$~pb & $2.4$~pb & $4.1$~pb & $4.0$~pb\\ 
 \% Higgs & 4.5\%/3.4\% & 4.2\%/6.8\% & 6.6\% & 12.8\% & 8.6\%  & 7.0\% \\     
 $\sigma_{h/H/A}$ & 0.18~pb/0.13~pb & 0.24~pb/0.55~pb & 0.18~pb & 0.31~pb & 0.35~pb & 0.28~pb \\
\hline   
 \end{tabular}
\caption{The parameters, part of the
  spectrum, and some relevant collider information 
 for the Study Points used in this analysis. 
 The spectrum was computed with SUSPECT2~\cite{Djouadi:2002ze}. The
 quoted cross section is determined at lowest order for the LHC
 operating at a center of mass energy of $\sqrt{s} = 14~\tev$. See the
 text for the definition of \% Higgs and $\sigma_{h/H/A}$.} 
\label{table:points}
\vspace{- 0 mm}
\end{table*}
\end{centering}

\subsection*{Backgrounds and Cuts:}

\noindent
The primary SM backgrounds we consider are:
\begin{itemize}
\item $\bar t t + $jets
\item $W/Z + $ heavy flavor
\item $W/Z +$ jets
\item $\bar t t + \bar b b$
\end{itemize}
These backgrounds are familiar from many supersymmetry/BSM
searches. They have large cross sections, multiple jets, some of which
are $b$-jets, and sources of missing energy from vector boson
decays. The background events are first generated at parton-level 
using ALPGEN~v13~\cite{Mangano:2002ea} and are then showered and
hadronized using PYTHIA~v6.4~\footnote{All events generated with
  ALPGEN using CTEQ5L parton distribution functions and default
  options for factorization/renormalization scheme.}.  We also use the  
ATLAS tune~\cite{Buttar:2004iy} in PYTHIA to model the underlying
event. Jet manipulation is done using FastJet~\cite{Cacciari:2005hq}.
We do not perform any detector simulation or smearing of jets. 
A realistic ATLAS/CMS specific search in the spirit of 
Ref.~\cite{atlas-study-vh} is beyond the scope of this work. 
However, since high $p_t$ jets result in a large amount of energy 
deposited in the calorimeter cells where energy resolution is
excellent, we do not expect smearing to significantly modify our
results.

Before we run our substructure algorithm, we introduce cuts to isolate
the signal from the background. Rather than tailoring the cuts to each
specific SHSP point, we choose a more generic set which can be applied
to all Study Points. In particular, we use: 
\begin{enumerate}
\item $\ETmiss > 300~\gev$.
\item $3^+$ jets, at least one of which is tagged as a b-jet. To be
  counted as a jet, we require $p_T > 200~\gev$ --  the $p_T$
  requirement on the jets is set so high because we want to capture an
  entire boosted object (ideally a Higgs) within a single jet. As
  explained in Sec.~\ref{sec:substructure}, objects will be
  reconstructed from subjets contained within individual high-$p_T$
  jets rather than combining multiple jets.  We impose a
  pseudorapidity cutoff of $|\eta| < 4.0$ for jets which are not
  flavor-tagged, while b-tagged jets are restricted by the
  pseudorapidity extent of the tracker, $|\eta| < 2.5$. 
\item No isolated leptons with $p_T > 20, |\eta_{\ell}| < 2.5$.
\item $\HTjets = \sum_{i} p_{T,i} > 1.0~\tev$, where the sum extends over
      all jets indexed by $i$.
\end{enumerate}
Large missing energy, large $\HTjets$, and high jet multiplicity are often the characteristics
of new physics and, in particular, of weak scale supersymmetry with
$R$-parity~\footnote{Due to our large jet-$p_T$ requirement,
  $\HTjets$ calculated with our definition can be quite different than
  the sum of all visible transverse energy in the calorimeters (often
  referred to as $H_{T, cal}$). However, $H_{T, cal}$, which we would
  rely on for triggering, will always be bigger than
  $\HTjets$.}. These variables are widely used in supersymmetric  
searches and we use them here. After $\ETmiss$ and $\HTjets$ cuts, 
the biggest background is by far $\bar t t + $jets.
In order to suppress the $\bar t t + $jets further we introduce a
lepton veto; the logic behind this cut is that any $\bar t t + $jets
events 
which pass the large $\ETmiss$ cut most likely contain at least one
leptonic $W^{\pm}$. 

We collect all events that pass our preliminary cuts and run the
substructure algorithm described in
Sec.~\ref{sec:substructure}. Events which pass the substructure
selection have at least one $b$-jet with substructure and,
consequently, at least one candidate resonance jet.   

The assumed background cross sections and their efficiencies under the
imposed cuts are summarized below in Table~\ref{table:sigbr}. To show
how substructure cuts affect the signal and background, we have broken
up the efficiencies into two stages. The first stage,
$\epsilon_{cuts}$, is calculated after the `conventional' cuts --
$\slashchar{E}_T, \HTjets$, jet multiplicity and lepton veto -- have
been imposed. Then, after running the substructure algorithm, the
surviving events are counted to determine $\epsilon_{cuts+subs}$.  

\begin{centering}
\begin{table}[h!]
\centering 
\begin{tabular}{ |c|c|c|c| }\hline
Process & $\sigma$ & $\epsilon_{cuts}$ & $\epsilon_{cuts + subs}$ \\ 
\hline \hline  
$\bar t t \,+ $ 0 jet & 474 pb & $0$& $0$ \\ 
$\bar t t \,+ $ 1 jet &  248 pb & $9.2\times 10^{-6}$& $1.62\times 10^{-6}$\\ 
$\bar t t \,+\ 2^+$ jet &  132 pb &$2.1\times 10^{-4}$& $4.84\times 10^{-5}$ \\  \hline
$\bar t t + \bar b b$ & 0.83 pb & $1.9\times 10^{-4}$& $4.6\times 10^{-5}$ \\ \hline
$W(\ell \nu) + 2\ \text{jets}$ & 127 pb & $2.3\times 10^{-6}$ & $0$\\
$W(\ell \nu) + 3^+\ \text{jets} $ & 50 pb & $2.3\times 10^{-4}$& $1.08\times 10^{-5}$ \\ \hline
$Z(\bar{\nu}\nu) + 2\ \text{jets}$ & 80 pb & $0$ & $0$ \\ 
$Z(\bar{\nu}\nu) + 3^+\ \text{jets}$ & 29 pb & $2.3\times 10^{-4}$& $1.47\times 10^{-5}$ \\ \hline
$Z(\bar{\nu}\nu) + \bar b b$ & 1.4 pb & $0$ & $0$  \\
$Z(\bar{\nu}\nu) + \bar b b + \text{jet}$ & 1.4 pb & $3.5\times 10^{-4}$ & $6.94\times 10^{-5}$\\ \hline
$W(\ell \nu) + \bar b b$ & 1.1 pb & $2.6\times 10^{-6}$ & $0$\\ 
$W(\ell \nu) + \bar b b + \text{jet}$ & 2.2 pb & $1.1\times 10^{-4}$& $3.6\times 10^{-5}$ \\ \hline \hline
SHSP 1A & 3.92 pb & 0.015 & $3.7\times 10^{-3}$\\ \hline
\end{tabular}
\caption{Signal 1A and background cross sections and efficiencies. The
  efficiency for the other Study Points is similar. All backgrounds
  were generated with parton level cuts on jets: $p_{T,j} > 30\ \gev$
  ($25\ \gev $ for $\bar t t$ + jets), $|\eta_j| < 4.0, \Delta R_{jj}
  > 0.4$. An additional cut $\slashchar E_T > 75\ \gev$ was used for
  all $W/Z$ backgrounds, while $|\eta_b|  < 2.75$ was added for all
  heavy-flavor events. All background cross sections are LO except for $t\bar t + \text{jets}$, which has been rescaled to the NLO MCFM~\cite{Nason:1987xz} result $\sigma = $\ 855 pb ($K \sim 1.8$).} 
\label{table:sigbr}
\end{table}
\end{centering}
As can be seen from the Table, the conventional cuts are quite
effective at reducing the background. The signal efficiency under the
conventional cuts looks low. However, as explained in
Sec.~{\ref{subsect:shsp}}, many supersymmetric events for these Study
Points come from electroweak chargino/neutralino pair production which
do not contain the sufficient energy or jet multiplicity to pass our
cuts; the efficiency for squark/gluino initiated events is
higher. Requiring jet substructure suppresses the background further
relative to the signal, however the real power from substructure comes
in the shape of the jet-mass distribution. Therefore, the final step
in our search strategy is to plot the invariant mass of all candidate
resonance jets and look for a peak consistent with a Higgs boson.  The
candidate resonance jet mass plots for each of the 8 benchmark Study
Points in Table~\ref{table:points} are presented in the following
sub-sections. To break up the results, we have grouped the Study
Points into the same three categories used in Table
\ref{table:points}: high-$m_A$, low-$m_A$, and one Study Point with a
LSP thermal relic density that matches cosmological observations.  
 
 In all of the following plots, the contribution from {\em all}
 supersymmetric events (inclusive superpartner production) are shown
 together on top of the SM background.  While the supersymmetric
 contribution contains our signal, Higgs bosons from sparticle decays,
 it also contains new backgrounds. Top quarks and $W/Z$ bosons will
 also be copiously produced in cascade decays and can occasionally
 pass the substructure cuts. In fact, in several circumstances this
 supersymmetric background is larger than the SM backgrounds. 
 
 \subsection{High-$m_A$ points: SHSP 1a/1b, 2a/2b}

\begin{figure}[!h]
\centering
\includegraphics[width=2.75in]{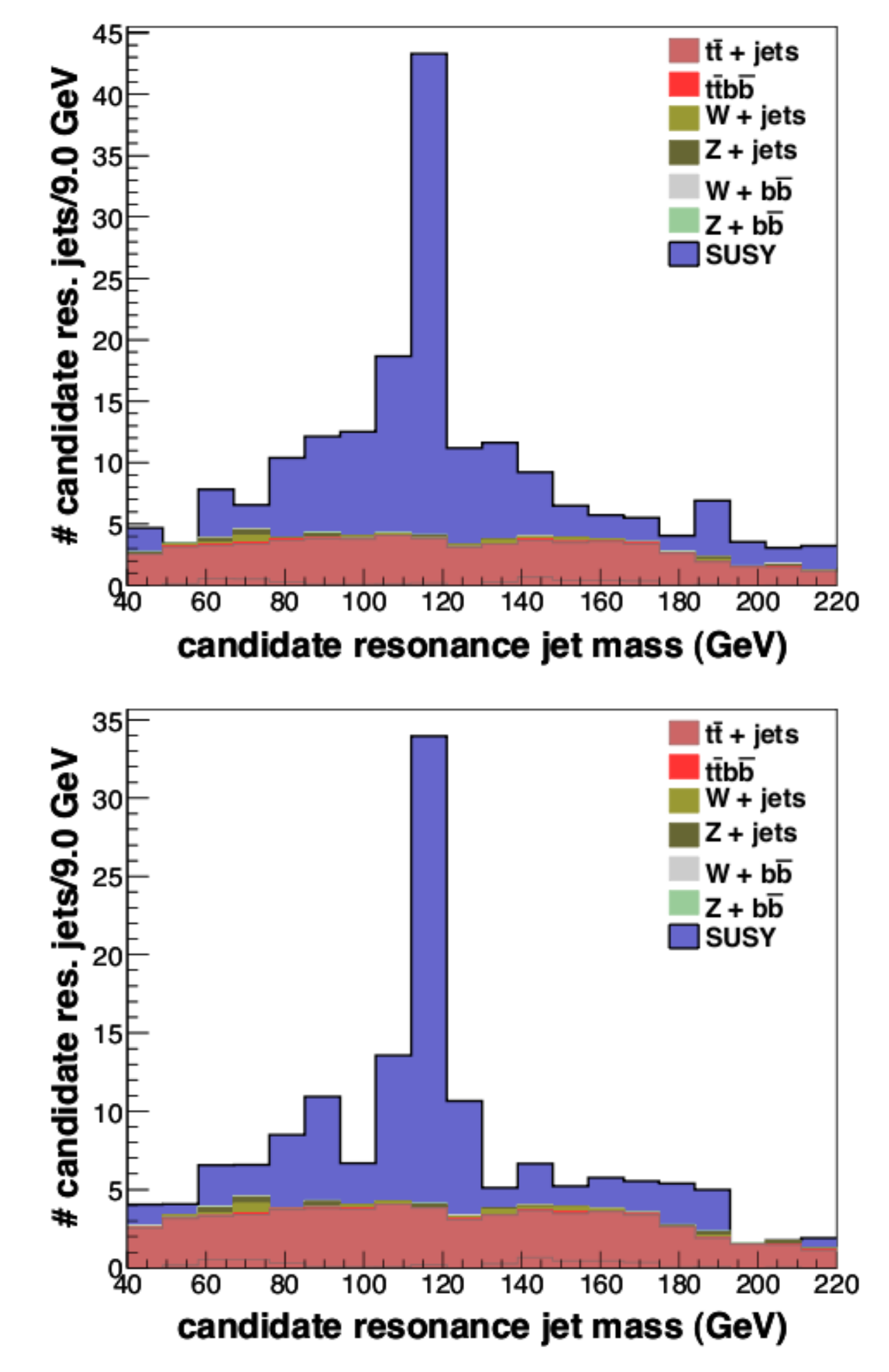}
\caption{Distribution of the candidate resonance jet mass normalized
  to $10\ \text{fb}^{-1}$ of integrated luminosity at $14\ \text{TeV}$
  center of mass energy. The contribution from supersymmetric
  particles is shown in blue for points SHSP~$1a$ (top) and SHSP~$1b$
  (bottom). Standard Model contributions, which come primarily from
  $t\bar t + \text{jets}$, are indicated by the red and green
  histograms.} 
\label{fig:shsp1}
\end{figure}

The first of the high-$m_A$ points, SHSP~$1a$ and~$1b$, are
characterized by small $\mu$. The large $m_A$ kinematically forbids
squark decays to other Higgs states  ($H$/$A$/$H^{\pm}$), while the
low $\mu = 0.5 M_1 = 150$~GeV implies a very Higgsino-like LSP and
thus large branching fractions $\chi^{\pm}_2 \rightarrow h +
\chi^{\pm}_1, \chi_{3,4} \rightarrow h + \chi_{1,2}$.  These points
are ideally suited to our analysis, and the resulting candidate
resonance jet mass plot, Fig.~\ref{fig:shsp1} verifies this. The peak
arising from Higgs decay is unmistakable over the relatively
featureless SM background. 

The small shoulder to the left of the Higgs peak comes from $Z
\rightarrow \bar b b$ events. Higgs bosons and the $Z$ are produced in
relatively equal amounts for these two points, due to the arguments
presented in Sec.~\ref{sec:goldstone}. However, the $Z \rightarrow
\bar b b$ branching fraction is only $1/6$ as large as  $h\rightarrow
\bar b b$, and the resulting $Z$ peak is small. 

The only difference between SHSP~$1a$ and $1b$ is the mass of the
sleptons. In SHSP~$1b$, the sleptons are light enough that the heavier
charginos and neutralinos can decay into them. As demonstrated in
Figs.~\ref{fig:br1a},\ref{fig:br1b},  new chargino/neutralino decay
modes imply a smaller fraction of decays to Higgs bosons, and thus a
smaller signal. However, comparing the top and bottom plots in
Fig.~\ref{fig:shsp1}, we can see that the rate decrease to due decays
to sleptons of mass $M_1 < m_{\tilde{l}} < M_2$ is quite minor. \\ 

To get a quantitative idea of how well our algorithm can find the
Higgs, we estimate the significance of the Higgs peak on top of the SM
and continuum new physics background. We determine the SM plus
continuum contribution using the same simple method as in
Ref.~\cite{Kribs:2009yh}; the histograms $1-2$ bins on either
side of the Higgs peak are connected with a line, and anything within
the resulting trapezoid is counted as background. To check the
veracity of this procedure, we have looked back into the signal events
and assigned each event with a candidate resonance jet to an initial
parent parton ($t/W/Z/h$) according to which heavy particle is closest
in $\Delta R$
We find the
fat-jets with a parent Higgs are indeed confined to the peak and
neighboring bins. Events with a $Z$ parent are similarly confined to
the bins near $m_Z$, while the continuum events are composed of $t/W$
events.  Using the peak $-1/+0$ bins to define the signal region
(meaning the bins $-2/+1$ on either side of the peak are used to
determine the background), we find $\mathcal S = S/\sqrt B \cong 7.9$ for point SHSP~$1a$. The same procedure, taking the signal region to be the peak $\pm 1$ bins, gives $\mathcal S = 9.6$ for point SHSP~$1b$. These significances are just rough estimates. We have taken quite aggressive conventional cuts to render the SM background as small and featureless as possible; less strict cuts may lead to higher significances, as would optimization of the cuts for each SHSP point.

The next two high-$m_A$ points are more challenging for three
reasons. First, points $2a$ and $2b$ have a smaller $M_1$. As we saw
in Figs.~\ref{fig:br1a},\ref{fig:br1b}, a lower $M_1/\mu$ means fewer
Higgs bosons are produced from squark cascades. Second, lowering $M_1$
while holding $M_1:M_2:M_3$ ratio fixed implies a much lighter
gluino. While the gluinos are light in this scenario, they are still
capable of decaying to on-shell squarks, so supersymmetric events
originating from gluinos -- either from gluino pair production or
squark-gluino associated production -- have more jets than events
originating from squark pairs. Additionally, because gluinos decay
democratically to all species of squarks, gluino cascades can easily
include top and bottom quarks.  The third difficulty with $2a$ and
$2b$ comes from right-handed squarks. Right-handed squarks, produced
either in pairs or associated with a left-handed squark or gluino,
decay to bino plus jet, with the bino in this spectra spread between
$\chi_1, \chi_2$, and  $\chi_3$. However, as can be seen from
Table.~\ref{table:sigbr}, the mass-gaps among the three lightest
neutralinos are  small enough so that 
most two-body decay modes are  kinematically forbidden; 
the neutralinos decay instead via an off-shell $W/Z/h$
plus a lighter chargino/neutralino. Off-shell, hadronic decays quickly
lead to an increase in the number of hard partons in the event. For
example, a typical signal process involving one right-handed squark:
$pp \rightarrow \tilde q_L \tilde q_R$ followed by $\tilde q_L
\rightarrow \chi^{0}_4 + j \rightarrow  \chi^0_1 + h +j$ and $\tilde
q_R \rightarrow \chi^0_3 + j \rightarrow \chi^0_1 + 3 j$ involves 4
extra hard partons, any one of which can fall in the same fat-jet as
the Higgs boson. Longer cascades, coming from gluino production or
more decay steps, are easy to imagine and will contain even more hard
partons. 

\begin{figure}[!h]
\centering
\includegraphics[width=2.75in]{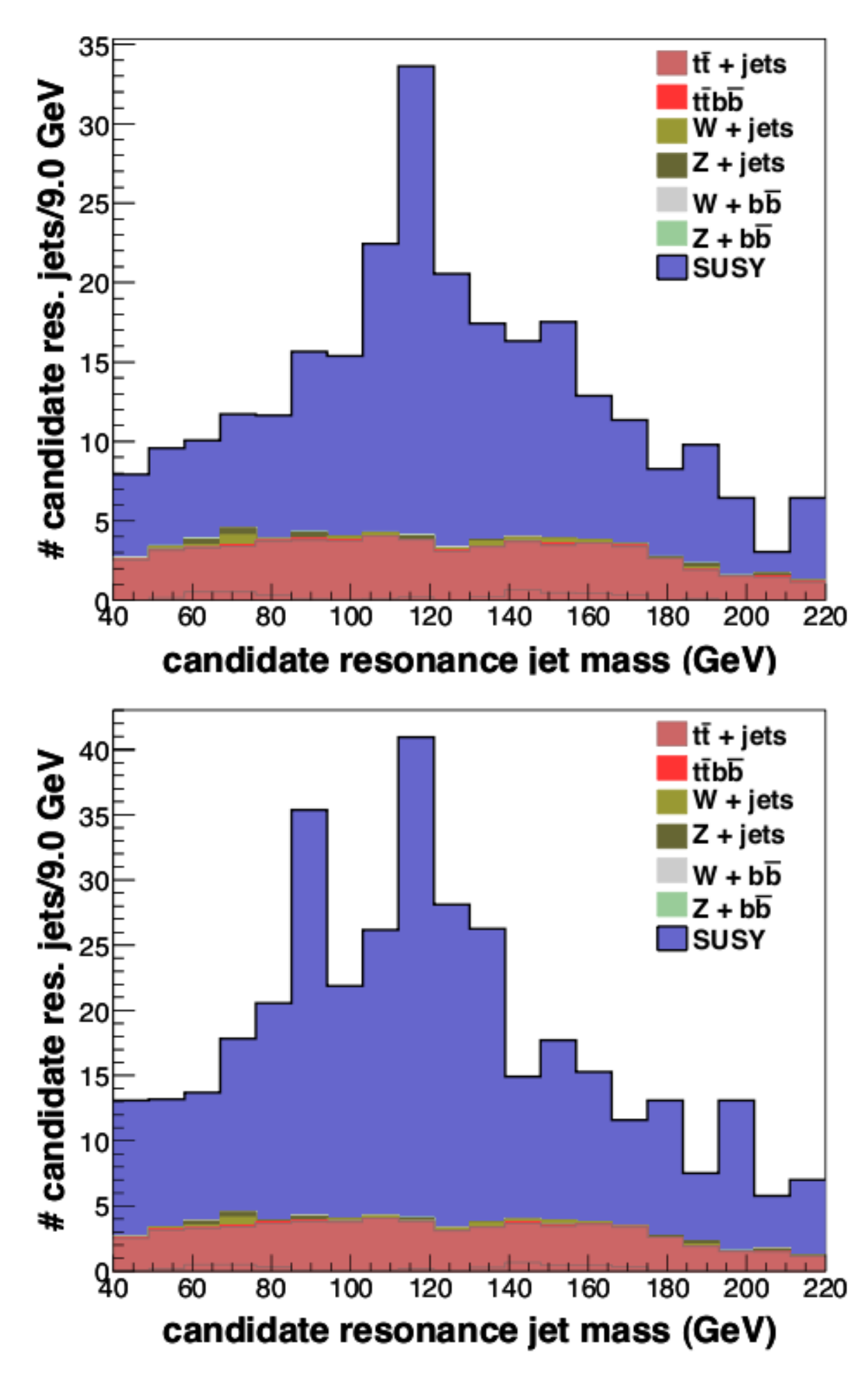}
\caption{Distribution of the candidate resonance jet mass in points
  SHSP 2a (top) and SHSP 2b (bottom). As in Fig.~\ref{fig:shsp1} we
  assume $10\ \text{fb}^{-1}$ of integrated luminosity and a $14\
  \text{TeV}$ center of mass energy.} 
\label{fig:shsp2}
\end{figure}

When extra partons from superpartner decays are erroneously combined
with all or part of a Higgs candidate, the jet mass becomes
smeared. The smearing is exacerbated by the fact that, following
BDRS, we take the three hardest subjets during filtering to 
define the candidate resonance jet. Such a procedure remarkably
improves the mass resolution of a Higgs jet when the correct threshold is
identified and none of the extra hard partons produced in association
with the Higgs is inside the Higgs cone. The three hard partons during
filtering then correctly capture  $b \bar b $ from Higgs as well as the
first radiation inside the $b \bar b$ system. On the other hand, if
there is an extra hard parton inside the  $b \bar b$ system, the
filtered resonance jet may end up containing $b \bar b \ + $ hard parton
instead of being $b \bar b \ + $ radiation and consequently  having a skewed
invariant mass. This smearing is clearly visible in
Fig.~\ref{fig:shsp2} and creates 
the feature extending from the Higgs peak to higher mass. However,
despite all the contamination from auxiliary hard partons, the Higgs
peak is still quite visible. Perhaps more elaborate subjet algorithms
could be used to clean up the high-mass tail further.

Moving from $2a$ to $2b$, the squark mass decreases. Lighter squarks
are produced even more prodigiously, as reflected in the enormous
superpartner cross section, however they impart a smaller boost on
their decay products. The increased production of right-handed squarks
in SHSP~$2b$,  a factor of $\sim 4$ compared to point $2a$ is
responsible for the increased number of supersymmetric events away
from the Higgs peak. The slepton mass in $2b$ is also smaller than in
$2a$, however this has only a small effect since the sleptons are
still too heavy for the higher-tier charginos and neutralinos to decay
into, $m_{\tilde l} > M_2$. Estimating significance in the same way as
we did for SHSP $1a/b$ and using $-1/+2$ bins to define the signal
region, we find a significance of $(3.8, 5.6)$ for points SHSP
($2a,2b$). 

Having seen the effects of decreasing the squark mass, it is natural
to ask what happens if we do the opposite and raise $m_{\tilde q}$ and
$M_3$ while keeping the rest of the supersymmetry parameters
fixed. The squark/gluino mass sets the scale for the boost of its
subsequent decay products, including any Higgs bosons. One may worry
that a higher sparticle scale would lead to Higgs decay products which
are so boosted  that  finite detector granularity or the need to b-tag
multiple subjets would render our algorithm useless. This does not
occur, however, as is evident in the distribution of the subjet
angular scale $R_{bb}$. We find a rather flat distribution between
$0.3 < R_{bb} < 1.2$, which persists even as $m_{\tilde q}/M_3$ is
raised to several TeV (squark with mass beyond $~3\ \tev$ have such
low production cross section that they become phenomenologically
irrelevant at the LHC). Therefore, even if granularity/tagging
inefficiencies ruin the most highly boosted Higgs bosons, the broad
tail of $R_{bb}$ indicates that our algorithm can remain viable
throughout the range of interesting squark masses. 

\subsection{Relic Point: SHSP 3}

The parameters of point SHSP 3 have been chosen such that the LSP has
a thermal relic abundance that matches cosmological observations for
the dark matter density. As described in Sec.~\ref{sec:relic}, this
requires delicately adjusting $M_1 \simeq |\mu|$ to get the right
bino/Higgsino admixture in the LSP. Point SHSP 3 also has light
sleptons, however the branching fraction to Higgs bosons is still high
enough for our analysis to be successful. The cascade of superpartner
decays contain more $Z$ than $h$, due to the smaller value of $M_1$,
but the Higgs peak remains clearly visible.  

\begin{figure}[!h]
\centering
\includegraphics[width=2.75in]{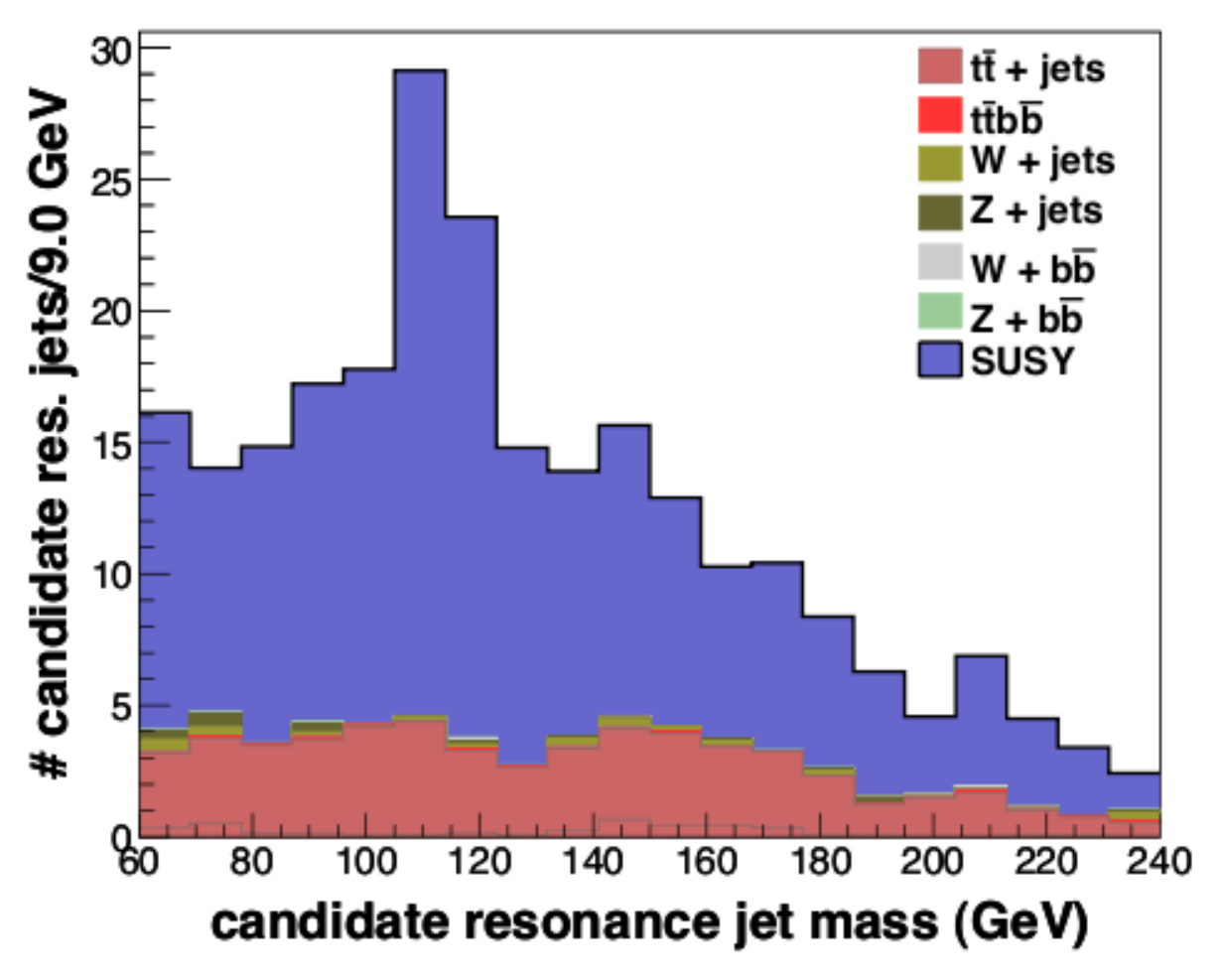}
\caption{Distribution of the candidate resonance jet mass in point
  SHSP 3. The parameters for this point were chosen to produce the
  correct dark matter relic abundance for the LSP. As in
  Fig.~\ref{fig:shsp1} we assume $10\ \text{fb}^{-1}$ of integrated
  luminosity and a $14\ \text{TeV}$ center of mass energy.} 
\end{figure}

\subsection{Low-$m_A$ points: SHSP 4-6}

The final set of Study Points have smaller $m_A$ and a small value for
$\tan{\beta}$. The region of smaller $m_A$, small $\tan{\beta}$ is
known to be difficult for traditional MSSM Higgs boson searches, so
these points serve as an important test of our algorithm. To ensure
the lightest Higgs boson has a mass that exceeds the LEP bound, we
allow larger mixing in the stop (and sbottom) sectors. These Study
Points are therefore quite similar to the ``maximal-mixing'' scenario
often considered in collider searches~\cite{Aglietti:2006ne}.   
 
\begin{figure}[!h]
\centering
\includegraphics[width=2.6in]{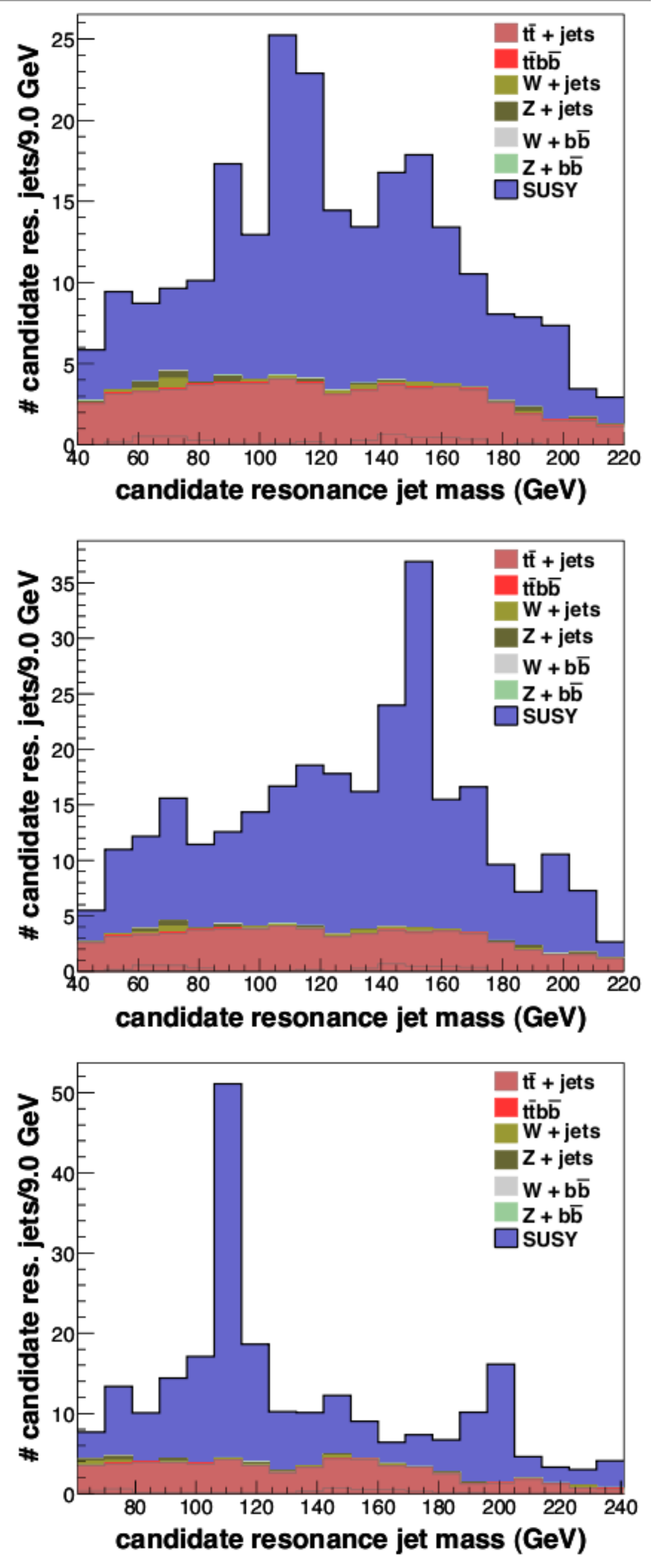}
caption{Distribution of the candidate resonance jet mass in points
  SHSP 4 (top), SHSP 5 (middle), SHSP 6 (bottom). As in
  Fig.~(\ref{fig:shsp1}) we assume $10\ \text{fb}^{-1}$ of integrated
  luminosity and a $14\ \text{TeV}$ center of mass energy.} 
\label{fig:shsp45}
\end{figure}

Perhaps the most interesting consequence of $m_A \ll m_{\tilde q},
M_2$, is that the heavier Higgs bosons $H/A$ also appear in the
superpartner decay cascades. For $m_H,m_A \simeq 150\ \gev$, the $H/A$
decay predominantly into $\bar b b$ and are light enough that they
will emerge from sparticle decays carrying a substantial boost. With
these characteristics, $H/A$ will be captured by our algorithm.  This
opens the exciting possibility, shown in fig. (\ref{fig:shsp45}), of
discovering multiple distinct Higgs bosons with a single analysis. 

In SHSP 4, the top plot of Fig.~\ref{fig:shsp45}, heavier charginos
and neutralinos decay to $h$ rather than $H/A$ making the $h$ peak
unmistakable. Some $H/A$ are present, and lead to the feature near
$m_A  = 150\ \gev$. Given the size of the $m_A$ feature and its
proximity to the top mass, detector resolution effects, which we have
treated very simply in this paper, become more important and need to
be taken into account correctly. $H/A$ discovery will likely require a
more specialized analysis, but it is certainly possible that both
$H/A$ and $h$ could be discovered with this technique given sufficient
integrated luminosity. 

 In point SHSP 5, the $\mu$ term is negative. With $\mu$ and $m_A$
 similar in magnitude, the Higgs mixing matrix becomes particularly
 sensitive to the relative sign between these two mass parameters and
 cancellations can occur once couplings are expressed in terms of mass
 eigenstates. For $\mu < 0$, the $h$ coupling to higher-tier
 charginos/neutralinos is suppressed by one such cancellation, and
 cascade decays to $H/A$ are more likely. We can clearly see this
 effect in fig.~(\ref{fig:shsp45}); the $h$ peak is barely visible
 over the continuum new physics events, while the narrow $H/A$ peak at
 $150\ \gev$ is clearly evident.   
 
 The final point, SHSP 6,  has exactly the same supersymmetry
 parameters as SHSP $1a$ except $m_A = 200\ \gev$. This is the ideal
 point for detecting both the light and heavy Higgs bosons with a
 single analysis. The $m_A$ is low enough that $\chi_4$ and
 $\chi^{\pm}_2$ have a moderate branching ratio to $H/A$, while $m_A$
 is heavy enough to avoid getting mistaken for new physics continuum
 or a top quark. Taking the signal region to be $-0/+1$ bins $(-1/+1)$
 around the $h$ peak, we find a significance of ($3.9,8.2$) for points
 (SHSP~$4$, SHSP~$6$). Repeating the same procedure around the $H/A$
 peak, we find a significance of  $(5.2, 4.5)$ for (SHSP~$5$,
 SHSP~$6$) using signal regions $-1/+0$ bin.

Low values for $m_A$ imply light charged Higgs bosons, which are
constrained by the flavor process $b \rightarrow s + \gamma$. While
the specific spectra we are looking at have $b\rightarrow s + \gamma$
slightly larger than the experimentally allowed
range~\cite{Djouadi:2002ze, HFAG}, slight changes in the spectrum,
such as lowering the third generation squark masses or introducing
squark mixing can introduce cancellations and significantly alter the
branching ratio $b\rightarrow s + \gamma$~\cite{Chen:2009cw}. These
changes to the spectrum need not effect the supersymmetric Higgs
signal. Therefore, in the same spirit as~\cite{Aad:2009wy,
  deRoeck:942733}, we 
focus on direct Higgs detection prospects and ignore indirect
constraints for the time being.

\section{Discussion}
\label{sec:discussion}

The power of using jet substructure with boosted Higgs decays 
into $b\bar{b}$ suggests the search for the MSSM Higgs bosons
should be entirely rethought and redone, with full detector
simulations.  Our estimates, without jet energy smearing
and without a realistic detector simulation, suggest 
that with less than $10$~fb$^{-1}$ of data at $\sqrt{s} = 14$~TeV, 
signal significance can exceed 5 for the $h \ra b\bar{b}$ channel 
alone given total superpartner production rate of order a few pb.  
This is possible
given the outstanding mass resolution of our reconstruction technique 
combined with the power that jet substructure provides in 
discriminating Standard Model and supersymmetric backgrounds.
We have been relatively conservative in our candidate resonance
jet finding algorithm given our flat $b$-tagging efficiency:  
we required a triple $b$-tag -- the original jet as well as two subjets.  
Nevertheless, our estimates of signal significance are just that --
estimates.  We urge the ATLAS and CMS collaborations to carry out
full detector simulations, along the lines of what was done
by ATLAS to study the boosted Higgs into $b\bar{b}$ mode 
in the Standard Model \cite{atlas-study-vh}.

The notion that both $h$ as well as $H$ and $A$ could be found
using jet substructure techniques is particularly interesting given
the difficulty that conventional search strategies have within
the smaller $m_A$ and smaller $\tan\beta$ region.  
The ATLAS and CMS TDR suggest fully covering the MSSM
parameter space requires considerable integrated luminosity,
$60$-$100$~fb$^{-1}$.  Our technique has the potential to
cover this region much more rapidly.  

It is interesting that the MSSM parameter region most favorable 
to finding a signal of Higgs bosons is also the one with the
least fine-tuning, namely, small $\mu$ (e.g.\ \cite{Kitano:2006ws}).
Nevertheless, gaugino mass unification, and other aspects of
the superpartner hierarchy are somewhat less constrained.

Finally, finding evidence for Higgs bosons within a new physics 
event sample provides an incredibly important connection between 
the new physics and the Higgs sector -- i.e., the Higgs sector 
is necessarily coupled with the new physics.
This connection can be established far faster than sorting out 
which kind of new physics is present based on the population
of different BSM search channels.
The generic search strategy proposed here builds on 
our previous paper \cite{Kribs:2009yh}, demonstrating the power of this 
method applied to the MSSM with a neutralino, or neutralino-equivalent, 
lightest supersymmetric particle.


\section*{Acknowledgments}

GDK thanks Fermilab and the Perimeter Institute and TSR thanks
Weizmann Institute and Fermilab for hospitality
where part of this work was completed.
This work was supported in part by the US Department of Energy 
under contract number DE-FG02-96ER40969 (GDK, TSR, MS).
AM is supported by Fermilab operated by Fermi Research Alliance, 
LLC under contract number DE-AC02-07CH11359 with the 
US Department of Energy.



\begin{thebibliography}{99}

\bibitem{Carena:2000dp}
  M.~S.~Carena, H.~E.~Haber, S.~Heinemeyer, W.~Hollik, C.~E.~M.~Wagner and G.~Weiglein,
  Nucl.\ Phys.\  B {\bf 580}, 29 (2000)
  [arXiv:hep-ph/0001002].

\bibitem{mssmlist}
  J.~F.~Gunion, P.~Kalyniak, M.~Soldate and P.~Galison,
  Phys.\ Rev.\  D {\bf 34}, 101 (1986);
  A.~Stange, W.~J.~Marciano and S.~Willenbrock,
  Phys.\ Rev.\  D {\bf 49}, 1354 (1994);
  J.~F.~Gunion, G.~L.~Kane and J.~Wudka,
  Nucl.\ Phys.\  B {\bf 299}, 231 (1988);
  J.~F.~Gunion and L.~H.~Orr,
  Phys.\ Rev.\  D {\bf 46}, 2052 (1992);
  J.~F.~Gunion and T.~Han,
  Phys.\ Rev.\  D {\bf 51}, 1051 (1995);
  R.~Kinnunen, S.~Lehti, A.~Nikitenko and P.~Salmi,
  J.\ Phys.\ G {\bf 31}, 71 (2005);
   J.~Dai, J.~F.~Gunion and R.~Vega,
  Phys.\ Lett.\  B {\bf 345}, 29 (1995);
  D.~Froidevaux and E.~Richter-Was,
  Z.\ Phys.\  C {\bf 67}, 213 (1995);
  M.~S.~Carena, S.~Mrenna and C.~E.~M.~Wagner,
  Phys.\ Rev.\  D {\bf 62}, 055008 (2000);
  C.~Balazs, J.~L.~Diaz-Cruz, H.~J.~He, T.~M.~P.~Tait and C.~P.~Yuan,
  Phys.\ Rev.\  D {\bf 59}, 055016 (1999);
  J.~L.~Diaz-Cruz, H.~J.~He, T.~M.~P.~Tait and C.~P.~Yuan,
  Phys.\ Rev.\ Lett.\  {\bf 80}, 4641 (1998);
  H.~S.~Hou, W.~G.~Ma, R.~Y.~Zhang, Y.~B.~Sun and P.~Wu,
  JHEP {\bf 0309}, 074 (2003);
  J.~j.~Cao, G.~p.~Gao, R.~J.~Oakes and J.~M.~Yang,
  Phys.\ Rev.\  D {\bf 68}, 075012 (2003);
  C.~Kao and N.~Stepanov,
  Phys.\ Rev.\  D {\bf 52}, 5025 (1995);
  A.~Belyaev, M.~Drees, O.~J.~P.~Eboli, J.~K.~Mizukoshi and S.~F.~Novaes,
  Phys.\ Rev.\  D {\bf 60}, 075008 (1999);
  G.~Cynolter, E.~Lendvai and G.~Pocsik,
  Acta Phys.\ Polon.\  B {\bf 31}, 1749 (2000);
  E.~Boos, A.~Djouadi and A.~Nikitenko,
  Phys.\ Lett.\  B {\bf 578}, 384 (2004);
  A.~Djouadi, W.~Kilian, M.~Muhlleitner and P.~M.~Zerwas,
  Eur.\ Phys.\ J.\  C {\bf 10}, 45 (1999);
  Z.~Kunszt and F.~Zwirner,
  Nucl.\ Phys.\  B {\bf 385}, 3 (1992);
  T.~Plehn, D.~L.~Rainwater and D.~Zeppenfeld,
  Phys.\ Lett.\  B {\bf 454}, 297 (1999).
 
\bibitem{Aad:2009wy}
  G.~Aad {\it et al.}  [The ATLAS Collaboration],
  arXiv:0901.0512 [hep-ex].
 
 
\bibitem{deRoeck:942733}
  A.~de Roeck, A.~Ball, M.~Della Negra, L.~Foa, and A.~Petrilli 
  [The CMS Collaboration],
  CERN-LHCC-2006-021 ; CMS-TDR-008-2.

\bibitem{Baer:1992ef}
  H.~Baer, M.~Bisset, X.~Tata and J.~Woodside,
  Phys.\ Rev.\  D {\bf 46}, 303 (1992).

\bibitem{Hinchliffe:1996iu}
 I.~Hinchliffe, F.~E.~Paige, M.~D.~Shapiro, J.~Soderqvist and W.~Yao,
 Phys.\ Rev.\  D {\bf 55}, 5520 (1997)
 [arXiv:hep-ph/9610544].

\bibitem{Datta:2001qs}
 A.~Datta, A.~Djouadi, M.~Guchait and Y.~Mambrini,
 Phys.\ Rev.\  D {\bf 65}, 015007 (2002)
 [arXiv:hep-ph/0107271];

\bibitem{Datta:2003iz}
 A.~Datta, A.~Djouadi, M.~Guchait and F.~Moortgat,
 Nucl.\ Phys.\  B {\bf 681}, 31 (2004)
 [arXiv:hep-ph/0303095].

\bibitem{Bandyopadhyay:2008fp}
 P.~Bandyopadhyay, A.~Datta and B.~Mukhopadhyaya,
 Phys.\ Lett.\  B {\bf 670}, 5 (2008)
 [arXiv:0806.2367 [hep-ph]].

\bibitem{Huitu:2008sa}
 K.~Huitu, R.~Kinnunen, J.~Laamanen, S.~Lehti, S.~Roy and T.~Salminen,
 Eur.\ Phys.\ J.\  C {\bf 58}, 591 (2008)
 [arXiv:0808.3094 [hep-ph]].

\bibitem{Bandyopadhyay:2008sd}
 P.~Bandyopadhyay,
 JHEP {\bf 0907}, 102 (2009)
 [arXiv:0811.2537 [hep-ph]].

\bibitem{Fowler:2009ay}
 A.~C.~Fowler and G.~Weiglein,
 JHEP {\bf 1001}, 108 (2010)
 [arXiv:0909.5165 [hep-ph]].
  
\bibitem{Butterworth:2008iy}
  J.~M.~Butterworth, A.~R.~Davison, M.~Rubin and G.~P.~Salam,
  Phys.\ Rev.\ Lett.\  {\bf 100}, 242001 (2008)
  [arXiv:0802.2470 [hep-ph]].

\bibitem{Kribs:2009yh}
  G.~D.~Kribs, A.~Martin, T.~S.~Roy {\it et al.},
  [arXiv:0912.4731 [hep-ph]].

\bibitem{Butterworth:2007ke}
  J.~M.~Butterworth, J.~R.~Ellis and A.~R.~Raklev,
  JHEP {\bf 0705}, 033 (2007)
  [arXiv:hep-ph/0702150].

\bibitem{Plehn:2009rk}
  T.~Plehn, G.~P.~Salam and M.~Spannowsky,
  arXiv:0910.5472 [hep-ph].

\bibitem{Martin:1997ns}
  S.~P.~Martin,
  In *Kane, G.L. (ed.): Perspectives on supersymmetry* 1-98.
  [hep-ph/9709356].

\bibitem{Djouadi:1997xx}
  A.~Djouadi, J.~L.~Kneur, G.~Moultaka,
  Phys.\ Rev.\ Lett.\  {\bf 80}, 1830-1833 (1998).
  [hep-ph/9711244].
  
\bibitem{Djouadi:1999dg}
  A.~Djouadi, J.~L.~Kneur, G.~Moultaka,
  Nucl.\ Phys.\  {\bf B569}, 53-81 (2000).
  [hep-ph/9903218].

\bibitem{atlas-study-vh}
  ATLAS Collaboration, ATL-PHYS-PUB-2009-088.
     
\bibitem{agashe}
K.~Agashe, A.~Belyaev, T.~Krupovnickas, G.~Perez and J.~Virzi,
 Phys.\ Rev.\  D {\bf 77}, 015003 (2008);

 \bibitem{gerbush}
 M.~Gerbush, T.~J.~Khoo, D.~J.~Phalen, A.~Pierce and D.~Tucker-Smith,
  Phys.\ Rev.\  D {\bf 77}, 095003 (2008);
 ATL-PHYS-CONF-2008-008 and ATL-COM-PHYS-2008-001, Feb.~2008

\bibitem{Brooijmans:2008zz}
  G.~Brooijmans,
  ATL-PHYS-CONF-2008-008, ATL-COM-PHYS-2008-001, Feb 2008.

\bibitem{Thaler:2008ju}
  J.~Thaler and L.~T.~Wang,
  JHEP {\bf 0807}, 092 (2008)
  [arXiv:0806.0023 [hep-ph]].

\bibitem{Kaplan:2008ie}
  D.~E.~Kaplan, K.~Rehermann, M.~D.~Schwartz and B.~Tweedie,
  Phys.\ Rev.\ Lett.\  {\bf 101}, 142001 (2008)
  [arXiv:0806.0848 [hep-ph]].

\bibitem{Almeida:2008yp}
  L.~G.~Almeida, S.~J.~Lee, G.~Perez, G.~Sterman, I.~Sung and J.~Virzi,
  Phys.\ Rev.\  D {\bf 79}, 074017 (2009)
  [arXiv:0807.0234 [hep-ph]].

\bibitem{Almeida:2008tp}
  L.~G.~Almeida, S.~J.~Lee, G.~Perez, I.~Sung and J.~Virzi,
  Phys.\ Rev.\  D {\bf 79}, 074012 (2009)
  [arXiv:0810.0934 [hep-ph]].
    
 \bibitem{Krohn:2009zg}
  D.~Krohn, J.~Thaler and L.~T.~Wang,
  JHEP {\bf 0906}, 059 (2009)
  [arXiv:0903.0392 [hep-ph]].
  
\bibitem{Ellis:2009su}
  S.~D.~Ellis, C.~K.~Vermilion and J.~R.~Walsh,
  Phys.\ Rev.\  D {\bf 80}, 051501 (2009)
  [arXiv:0903.5081 [hep-ph]].

\bibitem{Ellis:2009me}
  S.~D.~Ellis, C.~K.~Vermilion and J.~R.~Walsh,
  arXiv:0912.0033 [hep-ph].

\bibitem{Chekanov:2010vc}
 ÊS.~Chekanov and J.~Proudfoot,
 Ê
 ÊPhys.\ Rev.\ ÊD {\bf 81}, 114038 (2010)
 Ê[arXiv:1002.3982 [hep-ph]].
 Ê



\bibitem{Sjostrand:2006za}
  T.~Sjostrand, S.~Mrenna and P.~Skands,
  JHEP {\bf 0605}, 026 (2006)
  [arXiv:hep-ph/0603175].
  
\bibitem{Belanger:2010pz}
  G.~Belanger, F.~Boudjema, A.~Pukhov and A.~Semenov,
  arXiv:1005.4133 [hep-ph].
  
\bibitem{ArkaniHamed:2006mb}
  N.~Arkani-Hamed, A.~Delgado and G.~F.~Giudice,
  Nucl.\ Phys.\  B {\bf 741}, 108 (2006)
  [arXiv:hep-ph/0601041].


\bibitem{Butterworth:2009qa}
  J.~M.~Butterworth, J.~R.~Ellis, A.~R.~Raklev and G.~P.~Salam,
  Phys.\ Rev.\ Lett.\  {\bf 103}, 241803 (2009)
  [arXiv:0906.0728 [hep-ph]].

\bibitem{Krohn:2009th}
  D.~Krohn, J.~Thaler and L.~T.~Wang,
  JHEP {\bf 1002}, 084 (2010)
  [arXiv:0912.1342 [hep-ph]].
  
\bibitem{Soper:2010xk}
  D.~E.~Soper and M.~Spannowsky,
  arXiv:1005.0417 [hep-ph].

\bibitem{Hackstein:2010wk}
 ÊC.~Hackstein and M.~Spannowsky,
 Ê
 ÊarXiv:1008.2202 [hep-ph].
 Ê

\bibitem{Dokshitzer:1997in}
  Y.~L.~Dokshitzer, G.~D.~Leder, S.~Moretti and B.~R.~Webber,
  JHEP {\bf 9708}, 001 (1997)
  [arXiv:hep-ph/9707323].

\bibitem{Wobisch:1998wt}
  M.~Wobisch and T.~Wengler,
  arXiv:hep-ph/9907280.

\bibitem{Wobisch:2000dk}
  M.~Wobisch, PhD Thesis (2000).

\bibitem{Cacciari:2005hq}
  M.~Cacciari and G.~P.~Salam,
  Phys.\ Lett.\  B {\bf 641}, 57 (2006)
  [arXiv:hep-ph/0512210].

\bibitem{Bassetto:1984ik}
  A.~Bassetto, M.~Ciafaloni and G.~Marchesini,
  Phys.\ Rept.\  {\bf 100}, 201 (1983).

\bibitem{Djouadi:2002ze}
 A.~Djouadi, J.~L.~Kneur and G.~Moultaka,
 Comput.\ Phys.\ Commun.\  {\bf 176}, 426 (2007)
 [arXiv:hep-ph/0211331].

\bibitem{Mangano:2002ea}
  M.~L.~Mangano, M.~Moretti, F.~Piccinini, R.~Pittau and A.~D.~Polosa,
  JHEP {\bf 0307}, 001 (2003)
  [arXiv:hep-ph/0206293].

\bibitem{Buttar:2004iy}
  C.~M.~Buttar, D.~Clements, I.~Dawson and A.~Moraes,
  Acta Phys.\ Polon.\  B {\bf 35}, 433 (2004).
  
\bibitem{Nason:1987xz}
  P.~Nason, S.~Dawson, R.~K.~Ellis,
  Nucl.\ Phys.\  {\bf B303}, 607 (1988).
 
\bibitem{Aglietti:2006ne}
  U.~Aglietti, A.~Belyaev, S.~Berge {\it et al.},
    [hep-ph/0612172].

 \bibitem{HFAG}
R.~Bernhard et al [HFAG rare decay group],
http://www.slac.stanford.edu/xorg/hfag/rare/index.html

\bibitem{Chen:2009cw}
  N.~Chen, D.~Feldman, Z.~Liu {\it et al.},
  Phys.\ Lett.\  {\bf B685}, 174-181 (2010).
  [arXiv:0911.0217 [hep-ph]].

\bibitem{Kitano:2006ws}
  R.~Kitano and Y.~Nomura,
  arXiv:hep-ph/0606134.



\end{thebibliography}
\end{document}